\documentclass[11pt]{article}

\usepackage{amsmath}

\usepackage[english]{babel}
\usepackage[ansinew]{inputenc}
\usepackage{epsfig}
\usepackage{color,graphicx}
\usepackage{wrapfig}
\usepackage{amssymb,amsthm,amsmath}
\usepackage{dsfont}
\usepackage{natbib}
\usepackage{abstract}
\usepackage{enumerate}
\usepackage{comment}
\usepackage{fixmath}

\renewcommand{\abstractname}{Summary}

\newtheorem{theorem}{Theorem}[]
\newtheorem{lemma}{Lemma}[]

\newtheorem{proposition}{Proposition}[]

\newtheorem*{thma}{Proof}

\newcommand{\eqdef}{:=}

\newcommand\independent{\protect\mathpalette{\protect\independenT}{\perp}}
\def\independenT#1#2{\mathrel{\rlap{$#1#2$}\mkern2mu{#1#2}}}

\def\ds{\displaystyle}

\begin{document}


\title{\bf Exact Monte Carlo likelihood-based inference for jump-diffusion processes}
\author{\bf{F. B. Gon\c{c}alves$^{a,1}$, K. {\L}atuszy\'{n}ski$^{b,c}$, G. O. Roberts$^{b}$}}
\date{}

\maketitle

\begin{center}
{\footnotesize $^a$ Universidade Federal de Minas Gerais, Brazil\\
$^b$ University of Warwick, UK\\
$^c$ Alan Turing Institute}
\end{center}

\footnotetext[1]{Address: Av. Antonio Carlos, 6627 - DEST, ICEx, UFMG -
31270-901, Belo Horizonte, Minas Gerais, Brazil. E-mail: fbgoncalves@est.ufmg.br}


\renewcommand{\abstractname}{Abstract}
\begin{abstract}

Statistical inference for discretely observed jump-diffusion processes is a complex problem
which motivates new methodological challenges. Thus existing approaches invariably resort to time-discretisations which inevitably lead to approximations in inference.
In this paper, we give the first general collection of methodologies for exact (in this context meaning discretisation-free) likelihood-based inference for discretely observed finite activity jump-diffusions. The only sources of error involved are Monte Carlo error and convergence of EM or MCMC algorithms. We shall introduce both frequentist and Bayesian approaches, illustrating the methodology through simulated and real examples.


\end{abstract}


\section{Introduction}

Jump-diffusion processes are used in a variety of applications in several scientific areas, especially
economics \citep[see][]{cliff,darrell,Wolf,polson1,bjorn,johan,bns1,feng,kenn,MR2042661}.
Other applications can be found, for example, in
physics \citep[see][]{chud}, biomedicine \citep[see][]{GM94} and object
recognition \citep[see][]{GM02}.
Jump-diffusions are natural extensions to diffusions,  allowing for discrete discontinuities in trajectories which are otherwise described by diffusion dynamics, thus offering additional flexibility in modelling phenomena which exhibit sudden large jumps.

Inference for jump-diffusions is a challenging problem because transition densities are almost always intractable. This problem is typically overcome using approximations based on time-discretisations which typically lead to systematic biases which are difficult to quantify \citep[see][]{BL_Platen}. Therefore, the problem requires new approaches which circumvent the need for approximation, and we shall
call them \emph{exact} solutions. They are exact in the sense of not involving any kind of discrete-time approximation although other sources of inaccuracy are involved - Monte Carlo error and convergence of EM and MCMC algorithms, which are more reasonable to control. Whereas efficient exact solutions have already been proposed for the context where there is no jump component \citep[see][]{bpr06a,AoS,sermai}, the problem for jump-diffusions is extremely under-developed.

This paper provides a general suite of methodologies for exact likelihood-based inference for discretely observed finite activity jump-diffusions, featuring both maximum likelihood and Bayesian approaches. Some of our approaches directly use an algorithm that performs exact simulation of a class of jump-diffusion bridges proposed in \citet{flavio2} and called the \emph{Jump Bridge Exact Algorithm} (JBEA). However our most general approaches involve the construction of significant extensions of JBEA.

First we propose two methodologies (MCEM and MCMC) for the case where the drift and the jump-rate are uniformly bounded and JBEA can be directly applied. Our methodology collapses to that of \citet{bpr06a} in the case where no jump component is present. We also propose two methods for the general case where both the drift and the jump-rate may be unbounded. One of these is an importance sampling (IS) adaptation of the first MCEM method we propose providing improved Monte Carlo variance properties. The second method introduces a new perspective for solving the inference problem in an exact framework using an infinite-dimensional Markov chain Monte Carlo (MCMC) algorithm based on Barker's transitions \citep{Barker} and novel simulation techniques.

Our exact approach can be applied to a wide class of univariate models which allow
non-linear state-dependent drift and diffusion coefficients, and state-dependent and time-inhomogeneous jump rate and jump size distribution. The methodology extends in principle to certain multivariate jump-diffusions, although we do not explore this direction here.

Whilst the focus of this paper is clearly on exact inference for jump-diffusions, the use of
exact Barker's transitions via a {\em Two-Coin} algorithm which we will introduce
appears to be of generic interest for any context with intractable accept/reject ratios.
Barker's method is rarely used, as it is known to be uniformly dominated by the much more well-known Metropolis-Hastings  accept/reject formula. However, Metropolis-Hastings never beats Barker by a factor greater than $2$ in Peskun order sense \citep{LRB}, and crucially, in contrast to the Metropolis-Hastings algorithm, the particular form of the accept/reject formula in the Barker case permits the construction of the Two-Coin procedure.

This paper is organised as follows. The remainder of Section 1 provides a literature review and formally defines the class of jump-diffusion processes to be considered. Section 2 presents some technical background material used throughout the remainder of the paper. Section 3 presents two inference methods directly based on the exact simulation of jump-diffusion bridges and applicable when the drift and jump rates are uniformly bounded. Section 4 presents two other methods which make indirect use of the algorithm for exact simulation of jump-diffusion bridges and do not require the boundedness assumption. In Section 5
the methods are applied to simulated data sets to investigate their efficiency. Finally, two real data sets concerning the exchange rate GBP/USD  and the S\&P500 index are analysed in Section 6.

\subsection{Literature review on approximate methods}

Existing solutions for the inference problem that we consider are based on approximations which, in turn, rely on \emph{path discretisation} and/or \emph{data augmentation} strategies. These solutions follow basically three main directions: considering alternative estimators to the \emph{maximum likelihood estimator} (MLE); using numerical approximations to the unknown likelihood function; and estimating an approximation to the likelihood by using Monte Carlo methods. Moreover, several of the methods assume state-independence of various components of the model.

Alternative estimators can be found in \citet{dufie2} and \citet{dufie3}. Numerical approximations can be found in \citet{MR959611}, \citet{Ait} and \citet{Fili}. Particle filter-based Monte Carlo methods can be found in \citet{polson2} and \citet{polson3}. \citet{goligh} proposes a refinement where the particles are propagated via MCMC.
The author proposes a Metropolis-Hastings algorithm generalising the Durham and Gallant bridge (\cite{DG}) for the jump-diffusion case.

The most promising and robust solutions among the ones cited above rely on data augmentation, via Euler schemes, and Monte Carlo methods. An important drawback in data augmentation schemes is the fact that as more data points are input between the observations, so to reduce the error when using the discrete time approximation, the computational cost grows significantly. Moreover, \citet{BL_Platen} provide a result regarding the error of the strong Euler approximation, stating that a discrete time approximation $Y^{\Delta}$, with time step size $\Delta$, approximates a jump-diffusion $Y$ (satisfying some regularity conditions) at time $T$ such that
\begin{equation*}
\sqrt{E(|Y_T-Y_{T}^{\Delta}|^2)}\leq C\Delta^{0.5},
\end{equation*}
for $\Delta\in(0,\Delta_0)$, for a finite $\Delta_0>0$, where $C$ is a positive constant, independent of $\Delta$. Although the result provides the order (w.r.t. $\Delta$) of a bound on the error, no result about how to obtain $C$ is provided.

We emphasise that, whilst the methodologies proposed in this paper involve only Monte Carlo and EM or MCMC convergence errors and there exists a vast literature about how to control those type of errors, the most promising approximate solutions also have to deal with time discretisation errors. Furthermore, whilst improvement on the discrete-time error inevitably leads to an increase in computation cost, the exact methodologies proposed in this paper are free from such errors at a fixed and reasonable computational cost.

Recently, an exact methodology that performs unbiased Monte Carlo estimation of the transition density of a class of univariate jump-diffusions was proposed in \citet{giesecke2019simulated}, building on the work of \citet{pedersen}, \citet{AoS}, \citet{chen} and \citet{caserob}.
This work provides a methodology for exact MLE methods although suffers from two main drawbacks.
Firstly, the results that assure finite variance of the transition density estimator and consistency and asymptotic normality of the simulated MLE require strong assumptions that are typically hard to check and often not met in realistic situations. Secondly,
even when these  assumptions are met, the method can be very inefficient due to high importance sampling weight variances. Our methodology will circumvent these difficulties through the use of rejection sampling, importance sampling with provably finite variances, and a novel Barker's MCMC methodology for Bayesian inference.

\subsection{The jump-diffusion model}

Formally, a jump-diffusion is the stochastic process $V:=\{V_s:0\leq s
\leq t\}$ that solves the SDE:
\begin{equation}\label{e1}
\ds dV_s=b(V_{s-})ds+\sigma(V_{s-})dW_s+\int_Eg_1(z,V_{s-})m(dz,ds),\;\;V_0=v_0,
\end{equation}
where $b:\mathds{R}\rightarrow\mathds{R}$, $\sigma:\mathds{R}\rightarrow\mathds{R}^+$ and $g_1:E\times\mathds{R}\rightarrow\mathds{R}$ are assumed to satisfy the regularity conditions (locally Lipschitz, with a linear growth bound) to guarantee a unique weak solution \citep[see][Section 1.9]{platen}. Functions $b$ and $\sigma$ can be time-inhomogeneous, but we restrict ourselves to the time homogeneous case. $W_s$ is a Brownian motion and $m(dz,ds)$ is a random counting measure on the product space $E\times[0,t]$, for $E\subseteq\mathds{R}$, with associated intensity measure $\lambda_m$. We assume that $\lambda_m$ is absolutely continuous with respect to Lebesgue measure on $E\times[0,t]$ and Markov dependent on $V$:
\begin{equation}\label{lambda}
\ds \lambda_m(dz,ds;V_{s-})=\lambda_m(z,s;V_{s-})dzds=\lambda_1(s,V_{s-})f_Z(z;s)dzds,
\end{equation}
where, for any $v\in\mathds{R}$, $\lambda_1(\cdot,v)$ is a non-negative real valued function on $[0,t]$ and, for any $s\in[0,t]$, $f_Z(\cdot;s)$ is a standard density function with support $E$. According to (\ref{e1}) and (\ref{lambda}), between any two jumps, the process $V$ behaves as a homogeneous diffusion process with drift $b$ and diffusion coefficient $\sigma$. The jump times
follow a Markov point process on $[0,t]$ with intensity function (jump rate) $\lambda_1(s;V_{s-})$. A random variable $Z_j$ with density $f_Z(z;\tau_{j})$ is associated to each of the $N$ jump times $\tau_{1},\ldots,\tau_{N}$ and, along with the state of the process, determines the size of the jump $g_1(Z_j,V_{\tau_{j}-})$
at time $\tau_{j}$.


\section{Necessary technical background material}

\label{techmat}

Some of our methodology will rely  heavily on the
{\em Jump Bridge Exact Algorithm} for the exact simulation of jump-diffusion bridges (henceforth termed JBEA) introduced in  \citet{flavio2}.
JBEA simulates a finite-dimensional representation from the exact probability law of a class of univariate jump-diffusion bridges and can be used to derive methodologies to perform exact likelihood-based inference for discretely observed jump-diffusions. We refer to this finite-dimensional representation as a skeleton of the jump-diffusion bridge.

JBEA can only be directly applied to processes with unit diffusion coefficient. This is a genuine restriction in the multivariate case where not all diffusions can be reduced to this case. On the other hand, in the one-dimensional case, under weak regularity conditions we can always apply the Lamperti transform to obtain a process as required (assuming that the diffusion coefficient is continuously differentiable).
Thus we set $X_s=\eta(V_s)$ with the Lamperti transform:
\begin{equation}\label{transea}
\ds \eta(v)\eqdef \ds \int_{v^*}^{v}\frac{1}{\sigma(u)}du,
\end{equation}
where $v^*$ is some arbitrary element of the state space of $V$.

The transformed process $X:=\{X_s:0\leq s\leq t\}$ is a one-dimensional jump-diffusion solving the SDE:
\begin{equation}\label{ejd2}
\ds dX_s=\alpha(X_{s-})ds+dW_s+\int_Eg(z,X_{s-})m(dz,ds),\;\;\;\;X_0=\eta(v_0)=x_0,
\end{equation}
where the jump rate $\lambda(s,X_{s-})$ of $X$ is given by the original jump rate $\lambda_1$ applied at $\eta^{-1}(X_{s-})$ (the inverse of $\eta$). The expressions for $\alpha$ and $g$ are given in Appendix A, which also describes the details of JBEA.

\subsection{Some notational conventions}\label{subsec_not}

Consider the one-dimensional jump-diffusion process $V:=\{V_s:0\leq s\leq T\}$ solving the SDE in (\ref{e1}). Suppose that functions $b$, $\sigma$, $g_1$, $f_Z$ and $\lambda_1$ depend on an unknown parameter $\theta\in\Theta$ and that we observe $V$ at $(n+1)$ time instances $\mathbf{t}=(t_0,\ldots,t_n)$, with $t_0=0$ and $t_n=T$. We aim to carry out inference about the parameter vector $\theta$ based on the observations $\mathbf{v}=\{v_0,\ldots,v_n\}$ of $V$.

Note that $\eta$ and the functions $\alpha$, $\lambda$ and $g$, from the transformed process $X$, will also depend on $\theta$ and we shall introduce this dependence on the notation accordingly. Furthermore, we define $\mathbf{x}(\theta)=\{x_{0}(\theta),x_{1}(\theta),\ldots,x_{n}(\theta)\}$ as the transformed observations, which depend on $\theta$ whenever the diffusion coefficient $\sigma$ of $V$ does, and suppress the notation $(\theta)$ from $\mathbf{x}$ and $x_i$ whenever $\sigma$ does not depend on $\theta$.

From this point on, we shall also consider the following notation and definitions.
Let $\mathbb{P}$ be the probability measure of the jump-diffusion $X$ solving the SDE in (\ref{ejd2}) but with components (drift and jump process) indexed by the parameter vector $\theta$. Let $\ds \mathbb{Q}$ to be the measure of a jump-diffusion which has the same initial value as $\ds \mathbb{P}$ and is the sum
of a Brownian motion and a jump process with jump rate 1 and jump size density $f$ which does not depend on $\theta$ and such that $f_g\ll f$. We shall use the notations $\mathbb{P}$ and $\mathbb{Q}$ to refer to the original measures and all the respective measures induced by them upon (measurable) transformations and conditioning, which will be properly indicated in the brackets when writing, for example, $\ds \mathbb{P}(\cdot|\cdot)$.

Define $\Delta t_i=t_i-t_{i-1}$ and let $N_i$ be the number of jumps from $X$ in $(t_{i-1},t_i]$, $\tau_{i,j}$ and $J_{i,j}$ be the $j$-th jump time and jump size, respectively, for $j=1,\ldots,N_i$, and $\mathbold{\tau}$ be the set of all the jumps times in $[0,T]$. Let $f_g(\cdot;s,X_{s-},\theta)$ be the the jump size density of $X$ at time $s$. We shall use the abbreviated notation $\ell(i,j;\theta):=\lambda(\tau_{i,j},X_{\tau_{i,j}-};\theta)$ and $f_g(\cdot;i,j,\theta):=f_g(\cdot;\tau_{i,j},X_{\tau_{i,j}-};\theta)$. Finally, define $A(u;\theta):=\int_{0}^u\alpha(y;\theta)dy$ and $\Delta A(i,j;\theta):=A(X_{\tau_{i,j}};\theta)-A(X_{\tau_{i,j}-};\theta)$. We assume that $A(u;\theta)$ can be obtained analytically.

The inference algorithms to be presented in Section 3 require, among other things (see assumptions of the JBEA algorithm in Appendix A), that the function $\ds (\alpha^2+\alpha')$ is bounded below and functions $\alpha$ and $\lambda$ are bounded above. The algorithms from Section 4, however, do not require those two conditions and can then be applied to a much wider class of jump-diffusion processes. This greater applicability, however, comes at the price of a considerably higher computational complexity.

\section{Inference directly using JBEA}\label{JBEAalgs}

In this section we present two inference algorithms based directly on JBEA to perform exact inference for discretely observed jump-diffusion processes.
The first one is a Monte Carlo EM algorithm to find the MLE and the second one is a MCMC algorithm to sample from
the posterior distribution of the parameters.

\subsection{A Monte Carlo EM algorithm}\label{ssecmcem}

In this algorithm, JBEA is used to obtain a Monte Carlo estimate of the expectation on the E-step.
The exactness of the method is obtained by combining the exactness feature of JBEA with some auxiliary variable techniques.
We present two versions of the algorithm to cover the cases where the diffusion coefficient does or does not depend on unknown parameters.

\subsubsection{The case where the diffusion coefficient is known}\label{sssecmcem1}

In order to obtain the likelihood function of $\theta$ based on the observations $\textbf{x}$ we need the finite-dimensional distributions of $X$, which are typically unavailable.
We can, however, obtain the augmented or full likelihood function, that is obtained from observing the entire jump-diffusion trajectory in $[0,T]$.
This function is obtained by writing down the Radon-Nikodym derivative
of $\mathbb{P}$ w.r.t. any measure that dominates $\ds \mathbb{P}$ and $\ds \mathbb{Q}$ does not depend on $\theta$ \citep[see, for example,][]{GFlik}.
This is the natural environment to apply the \emph{EM algorithm}.
Note, however, that the missing data is an infinite-dimensional random variable in this case and, for that reason, we have to be particularly careful in constructing a non-degenerate algorithm.

We define $X_{mis}$ as the unobserved part of the process $X$ consisting of the bridges of $X$ between the observations $\textbf{x}$.
We use $\ds \mathbb{Q}$ as the dominating measure to obtain the likelihood of a complete path of $X$. Theorem 2 from \citet{flavio2} implies that the complete log-likelihood function is given by
\begin{eqnarray}\label{llik}
l(X|\theta)&=&A(x_n;\theta)-A(x_0;\theta)-\sum_{i=1}^{n}\left(\sum_{j=1}^{N_i}\left(\Delta A(i,j;\theta)\right)-\int_{t_{i-1}}^{t_i}\phi(s,X_{s};\theta)ds\right) \nonumber \\
&&\ds +\ds\sum_{i=1}^{n}\sum_{j=1}^{N_i}\log\left(\ell(i,j;\theta)\right)+\log\left(f_{g}(J_{i,j};i,j,\theta)\right)+\kappa,
\end{eqnarray}
where \begin{equation}\label{phi_def}
\ds \phi(s,X_{s};\theta)=\left(\frac{\alpha^2+\alpha'}{2}\right)(X_{s};\theta)+\lambda(s,X_{s};\theta)
\end{equation}
and
$\kappa$ is a constant with respect to $\theta$ and can, therefore, be neglected in the EM algorithm.
Since the expectation of (\ref{llik}) cannot be evaluated analytically, we rely on Monte Carlo methods to
obtain an unbiased and strongly consistent estimator of it.

Given the auxiliary variable $U=(U_1,\ldots,U_n)$, where the $U_i$'s are mutually independent with $U_i\sim U(t_{i-1},t_i)$ and independent of $X$, we have that:
\begin{eqnarray}\label{expc2}
\ds && \mathbb{E}_{X_{mis}|\mathbf{x};\theta'}\left[l(X|\theta)\right]=A(x_n;\theta)-A(x_0;\theta) - \mathbb{E}\left[\sum_{i=1}^{n}\sum_{j=1}^{N_i}\left(\Delta A(i,j;\theta);\theta)\right)\right] +\mathbb{E}(\kappa) \nonumber \\
&&-\mathbb{E}\left[\sum_{i=1}^n\Delta t_i\ds\phi(U_i,X_{U_i};\theta)\right] 
+\mathbb{E}\left[\sum_{i=1}^{n}\sum_{j=1}^{N_i}\log\left(\ell(i,j;\theta)\right)+\log\left(f_{g}(J_{i,j};i,j,\theta)\right)\right],
\label{MCEMest}
\end{eqnarray}
with the subscript $(X_{mis},U|\mathbf{x};\theta')$ being omitted from all the expectations on the r.h.s. of (\ref{expc2}).

On the E-step, an estimate of the r.h.s. in (\ref{expc2}) is obtained via simple Monte Carlo integration, based on $M$ iid samples from $(X_{mis},U|\mathbf{x};\theta')$. Due to the Markov property of $X$, each one of the $M$ samples is independently obtained from the $n$ bridges of $\ds X\sim\mathbb{P}$ in intervals $(t_{i-1},t_i)$, conditional on the respective values of $x_{i-1}$ and $x_i$, for $i=1,\ldots,n$, via JBEA. This also includes the simulation of $X$ at times $U$. On the M-step, the estimate obtained on the E-step is maximised w.r.t. $\theta$. The value obtained from this maximisation becomes the new $\theta'$ for a new iteration of the algorithm.
Thus a single iteration of the algorithm inputs $\theta'$ and outputs the maximising $\theta$ value. The algorithm iterates this procedure until the output values are convergent according to the desired accuracy.

It is well documented \citep[see for example][and references therein]{Fort} that the number of Monte Carlo samples should increase with the EM iterations
in order to overcome Monte Carlo error.
Finally, the maximisation step may require numerical methods. For well behaved likelihoods, standard numerical optimisation algorithms should work well. For example, the quasi-Newton method BFGS \citep[see, for example,][Section 3.2]{BFGS} is used in the example presented in Subsection \ref{ss51}.

\subsubsection{The case where the diffusion coefficient is unknown}\label{MCEMudc}

We now focus on the case where the diffusion coefficient of $V$ depends on
unknown parameters.
A relevant result in (jump-)diffusion theory is that a complete path of a (jump-)diffusion can be used to almost surely perfectly estimate
$\ds \sigma(V_{s-};\theta)$. The result for the jump-diffusion case states that \citep[][II.6]{protter}:
\begin{equation}\label{qvt2}
\ds \lim_{n\rightarrow\infty}\sum_{l=1}^n\left(V_{\frac{lT}{n}}-V_{\frac{(l-1)T}{n}}\right)^2=\int_{0}^T\sigma^2(V_{s};\theta)ds+\sum_{j=1}^{N}(V_{\tau_{j}} -  V_{\tau_{j}-})^2.
\end{equation}

The computational implication of this result is that we cannot
construct an EM algorithm as in Subsection \ref{sssecmcem1} because there is a perfect correlation between $\sigma $ and the missing path as described in (\ref{qvt2}) \citep[see][]{rubinmeng}.
This problem is also relevant in MCMC algorithms, where it was first encountered in the context of inference for diffusions \citep[see][]{Stramer,Elerian}.

We propose a solution for this problem which combines together and then substantially generalises  \citet{Stramer} and \citet{bpr06a}. The method consists
of a suitable transformation of the missing data that breaks the dependence between the missing data and the parameters,
when conditioned on the observed values.

In our infinite-dimensional context, this problem is equivalent to finding a
reparameterisation of the missing data so that the dominating measure is independent of the parameters.
We construct this reparameterisation in two easily interpreted transformations.

The first transformation considers the Lamperti transformed process $X=\eta(V;\theta)$ as described in (\ref{transea}).
We make $\tau_{i,0}=t_{i-1}$ and $\tau_{i,{N_i+1}}=t_i$ to obtain the second level of path transformation $\ds \{X_{s}\rightarrow\dot{X}_{s};\;s\in(t_{i-1},t_i)\setminus\{\tau_{i,1},\ldots,\tau_{i,{N_i}}\}\}$, for $i=1,\ldots,n$ and $j=1,\ldots,N_i+1$:
\begin{equation}\label{trans2}
\ds    \dot{X}_{s}:= X_{s}-\left(1-\frac{s-\tau_{i,{j-1}}}{\tau_{i,j}-\tau_{i,{j-1}}}\right)X_{\tau_{i,j-1}}-
\left(\frac{s-\tau_{i,{j-1}}}{\tau_{i,j}-\tau_{i,{j-1}}}\right)X_{\tau_{i,j}-};\;s\in(\tau_{i,{j-1}},\tau_{i,j}),
\end{equation}
where $\ds X_{\tau_{i,0}}=x_{i-1}(\theta)$ and $\ds X_{\tau_{i,N_i+1}-}=x_{i}(\theta)$.

Note that $\dot{X}:=\{X_s;s\in[0,T] \} \setminus \{\mathbf{t},\mathbold{\tau}\}$ is a collection of diffusion bridges starting and ending at 0 between the observation and jump times. Its dynamics depend on $\theta$ and
are typically intractable; nevertheless it is easy to simulate $\dot{X}$ at any time $s$, conditionally
on $\mathbf{v}$ and a specific value of $\theta$, by firstly computing $\ds x_{i-1}(\theta)$ and $\ds x_{i}(\theta)$, then simulating $X$ conditioned on these two values via JBEA and, finally, applying the transformation in (\ref{trans2}). This will allow us to use a $\theta$-free dominating measure to obtain the likelihood function of a complete path of $X$ (see the proof of Lemma \ref{comlliklem} in Appendix C). Furthermore, the MCEM algorithm requires the simulation of one bridge point of $\dot{X}$ at a uniformly chosen time instant $U_i$ in every interval $(t_{i-1},t_i)$, as it is shown further ahead.

We define $\mathbf{x}(\theta)$ as the vector of the transformed observations. The inverse transformation to obtain $X_{s}$ from $\dot{X}_{s}$ is given by
\begin{equation}\label{trans3}
\ds    X_{s}=\varphi(s,V_{mis},\mathbf{x}(\theta))=\dot{X}_{s}+\left(1-\frac{s-\tau_{i,{j-1}}}{\tau_{i,j}-\tau_{i,{j-1}}}\right)X_{i,j-1}+
\left(\frac{s-\tau_{i,{j-1}}}{\tau_{i,j}-\tau_{i,{j-1}}}\right)X_{i,j-};\;s\in(\tau_{i,{j-1}},\tau_{i,j}).
\end{equation}
The data augmentation scheme is now based on $V_{mis}=(\mathbf{J},X_{J},\dot{X})$, where $\mathbf{J}$ is the jump process (jump times and sizes) of $X$ and $X_J$ is $X$ at its jump times. We define $V_{com}=(\textbf{v},V_{mis})$.

We obtain the likelihood function of $\theta$ by writing the joint law of $\mathbf{v}$ and
a suitable transformation of $V_{mis}$ (see the proof of Lemma \ref{comlliklem} in appendix C) with respect
to a dominating measure that does not depend on $\theta$. First, define
\begin{equation}\label{phidot_def}
\ds \dot{\phi}(s,\dot{X}_{s};\theta):=\phi(s,\varphi(s,V_{mis},\mathbf{x}(\theta));\theta).
\end{equation}

\begin{lemma}\label{comlliklem}
The likelihood of $\theta$ given the complete data $V_{com}$ is given by
\begin{eqnarray}\label{comllik}
\ds    \exp\{l(V_{com}|\theta)\}&\propto& \exp\left\{A(x_n(\theta);\theta)-A(x_0(\theta);\theta)-\sum_{i=1}^{n}\sum_{j=1}^{N_i}\Delta A(i,j;\theta) \right\} \nonumber \\
&\times&\ds\exp\left\{ - \sum_{i=1}^{n}\int_{t_{i-1}}^{t_i}\dot{\phi}(s,\dot{X}_{s};\theta)ds\right\} \prod_{i=1}^{n}\prod_{j=1}^{N_i}\left[\ell(i,j;\theta)f_{g}(J_{i,j};i,j,\theta)\right] \nonumber \\
&\times&\ds\prod_{i=1}^{n}\left[|\sigma(v_i;\theta)^{-1}| f_{N}(X_{i,1-};x_{i-1}(\theta),\tau_{i,1}-t_{i-1}) f_{N}(x_i(\theta);X_{i,N_i},t_i-\tau_{i,{N_i}}) \right],
\end{eqnarray}
where $f_N(u;a,b)$ is the Lebesgue density of the normal distribution with mean $a$ and variance $b$ evaluated at $u$.
\end{lemma}
\begin{thma}
See Appendix C.
\end{thma}
\noindent Multiplicative terms in $\exp\{l(V_{com}|\theta)\}$ that do not depend on $\theta$, including some normal p.d.f.'s, are omitted in expression (\ref{comllik}).

The algorithm is now analogous to the case where $\sigma$ is known. On the E-step, $M$ iid samples of $(V_{mis},U|\mathbf{v};\theta')$ are obtained via JBEA (and applying the transformation in (\ref{trans2}) to obtain $\dot{X}$) and used to compute a Monte Carlo estimate of the expectation of the log of (\ref{comllik}) w.r.t. the measure of $(V_{mis},U|\mathbf{v};\theta')$.
Notice that\\
$\ds \mathbb{E}\left[ \sum_{i=1}^{n}\int_{t_{i-1}}^{t_i}\dot{\phi}(s,\dot{X}_{s};\theta)ds \right] = \mathbb{E}\left[\sum_{i=1}^n\Delta t_i\ds\dot{\phi}(U_i,\dot{X}_{U_i};\theta)\right]$, where the expectation on the l.h.s. is w.r.t. $(V_{mis}|\mathbf{v};\theta')$ and the one on the r.h.s. is w.r.t. $(V_{mis},U|\mathbf{v};\theta')$.
On the M-step, the estimate obtained on the E-step is maximised w.r.t. $\theta$ and the obtained value becomes the new $\theta'$ for a new iteration of the algorithm.
Care must be exercised to keep track of all terms that depend on $\theta$, for example, $\varphi(\cdot,V_{mis},\mathbf{x}(\theta))$ and the $x_{i}(\theta)$'s.

\subsection{A Markov chain Monte Carlo approach}\label{MCMC_1}

We now present a Bayesian solution for the inference problem with parameter-dependent diffusion coefficient by
constructing a Markov chain with stationary distribution given by the exact joint posterior distribution of $\theta$ and the variables output by JBEA.
More specifically we have a standard Metropolis-Hastings (MH) algorithm to updated $\theta$, interspersed by moves to update a skeleton ${\bf S}$ of the latent jump-diffusion bridges of $X$, as introduced in Section \ref{techmat} and described in Appendix A, in $(0,T)\setminus {\bf t}$, where ${\bf S}$ is the union of skeletons $S^{(i)}$, for all $i$.

In theory, to be exact, our MCMC scheme will require the imputation and storage of entire continuous-time trajectories of jump-diffusion bridges between observations. The key to the retrospective simulation approach is to note that subsequent parameter updates within our MCMC scheme can be carried out exactly by only requiring a finite
but random collection of imputed bridge values (stored within a skeleton).
The skeletons $S^{(i)}$ are sampled independently, by applying JBEA to each of the intervals $(t_{i-1},t_i)$. The conditional independence of $S^{(i)}$'s, given ${\bf x}$ and $\theta$, is implied by the Markov property. A skeleton $S^{(i)}$ contains
\begin{enumerate}
\item
the jump times and sizes of $X$ in $(t_{i-1},t_i]$,
\item a random collection $\Phi^{(i)}$ of points in the rectangle $[t_{i-1},t_i]\times[0,1]$,
\item
the value of $X$ at the jump times and at the times given by the horizontal coordinates of $\Phi^{(i)}$, and
\item
a finite-dimensional measurable function ${\textbf L_{i}}$ of $X$, termed its {\em layer}, on the interval $[t_{i-1},t_i]$ (and motivated and described below).
\end{enumerate}
In order for the subsequent parameters update steps within our MCMC procedure to be carried out, it turns out that we need at least to know finite upper and lower bounds on $X$ within $[t_{i-1},t_i]$.  ${\textbf L_{i}}$ therefore needs to facilitate these bounds. However, since skeletons are dynamic (we need to be able to add extra time points at which $X$ is evaluated) it is also vital that we can carry out conditional simulation of such extra points conditional on ${\textbf L_{i}}$. A formulation for ${\textbf L_{i}}$ which satisfies these two conditions was presented in Appendix E, based on the work of \citet{bpr07}. We give an informal description here, while referring the interested reader to Appendix E for full details.

Writing
$$
{\textbf L_{i}} = (L_{i,1}, L_{i,2}, \ldots , L_{i,N_i+1})
$$
and given a sequence of positive increasing thresholds $\{b_k;\;k=0, 1, 2, \ldots \}$, with $b_0 = 0$, we define each component
$$
L_{i,j}=1+\sup \left\{k; \sup_{s\in (\tau_{i,j-1}-\tau_{i,j}) }|\dot{X}_s| \ge b_k\right\}\ .
$$
This way, $L_{i,j}=K$ implies that $\dot{X}_s\in(-b_K,b_K)$, for $s\in (\tau_{i,j-1},\tau_{i,j})$.

Let $\pi(\theta)$ be the prior density of $\theta$. We shall use $\pi$ as a general notation for densities. It is likely that we are primarily interested in the posterior distribution $\pi(\theta|\textbf{v})$ of $\theta$ although, depending on the application, we might also be interested in the posterior distribution $\pi({\bf J},X_J,\dot{X}|\textbf{v})$, which can be obtained from the joint posterior $\ds \pi\left(\theta,{\bf S}|\textbf{v}\right)$.

The full conditional density of $\theta$ is given by Theorem 1 in Appendix B. It may be convenient to sample $\theta$ in blocks as
it may be possible to simulate directly from the full conditional distributions of parameters in the jump rate and jump size distribution when conjugated priors are used.
This way a MH step is only used for parameters in the drift and diffusion coefficient.

\section{Inference when exact simulation of bridges is not possible}\label{inf2}

The two algorithms proposed in Section \ref{JBEAalgs} require JBEA to be directly applied, meaning that the drift $\alpha$ and the jump rate need to be bounded and function $(\alpha^2+\alpha')$ need to be bounded below. In order to circumvent those restrictions, we present two algorithms to perform exact inference which do not make direct use of JBEA, and, therefore, can be applied to a more general class of jump-diffusion models. Formally, the two algorithms only require assumptions \emph{(a)}, \emph{(c)}, \emph{(e)} and from Appendix A to be satisfied and that $A(u)$ can be obtained analytically. One of the key points in the proposed algorithms is the use of local bounds for the proposed path, which are obtained from the simulation of functions ${\textbf L_{i}}$, as introduced in Subsection \ref{MCMC_1}, through an algorithm called the layered Brownian bridge (see Appendix E). These bounds allow us to obtain local lower and upper bounds for the function $\left(\frac{\alpha^2+\alpha'}{2}+\lambda\right)$, as required by the algorithms in this section.

The first approach introduces an MCEM algorithm that follows similar lines to that previously presented
in Subsection \ref{ssecmcem} but uses an importance sampling estimate
for the E-step. The second approach introduces an infinite-dimensional MCMC algorithm which is
quite different from that in Subsection \ref{MCMC_1}.

\subsection{Importance Sampling MCEM}\label{ssecismcem}

For simplicity, we present here the algorithm for the case in which the diffusion coefficient is parameter-free. The algorithm for the general case can be obtained by combining results from Subsection \ref{MCEMudc} with this.

The algorithm requires that we can find a dominating measure from which we can simulate and w.r.t. which the measure of bridges of $\ds X\sim\mathbb{P}$
is absolutely continuous. However, unlike the algorithm from Subsection \ref{ssecmcem}, the implied RN derivative is not required to be uniformly bounded.
The algorithm proposed in this section outperforms the algorithm from Subsection \ref{ssecmcem}, when the latter is feasible and given that the proposal measure from which the samples are drawn are the same in both algorithms, in the sense of having a smaller variance MC estimator of the expectation on the E-step (it is well known in the literature that importance sampling outperforms rejection sampling in terms of MC variance). On the other hand, in our context, the IS estimator has a higher computational cost, given the extra variables and calculations involved due to the intractability of the original weights (to be detailed further ahead).

We introduce the probability measure $\mathbb{D}$ that, restricted to the interval $(t_{i-1},t_i)$, is the law of a bridge process conditioned to start at $x_{i-1}$ and end at $x_i$, defined as the sum of a jump process and a continuous process. The former is, marginally, a jump process with a constant positive jump rate $\lambda_i(\theta)$ and jump size density $f_i(\cdot;\theta)$ and the latter is, conditional on the former, a Brownian bridge that starts at $x_{i-1}$ at time $t_{i-1}$ and finishes in $x_i-J_i$ at time $t_i$, where $J_i$ is the sum of the jumps of the jump process. The jump components $\lambda_i(\theta)$ and $f_i(\cdot;\theta)$ are assumed to be time- and state-independent, although the extension to make them time dependent is straightforward \citep[see][]{flavio2}. We also assume $\mathbb{D}$ to be mutually independent among the intervals $(t_{i-1},t_i)$. In order to sample from $\mathbb{D}$ in $(t_{i-1},t_i)$ we use the following algorithm.\\
\\
\begin{tabular}[!]{|l|}
\hline\\
\parbox[!]{17cm}{
\texttt{
{\bf Algorithm 1: sampling from $\mathbb{D}$ in $(t_{i-1},t_i)$} {\scriptsize
\begin{enumerate}\label{2ca}
\setlength\itemsep{-0.5em}
  \item Sample the jump process with jump rate $\lambda_i(\theta)$ and jump size density $f_i(\cdot;\theta)$;
  \item sample the value of the process at the jump times by sampling a Brownian bridge (starting at $x_{i-1}$ at time 0 and ending at $(x_i-J_i)$ at time $\Delta t_i$) at the jump times and adding the jumps;
  \item sample the standard Brownian bridges $\dot{X}_s$ between times $\{t_{i-1},\tau_1,\ldots,\tau_{N_i},t_i\}$ any required time instant;
  \item obtain the value of $X(s)$ at the required time instants mentioned on the previous step by applying the transformation $\varphi$ as defined in (\ref{trans3}).
\end{enumerate}}
}}\\  \hline
\end{tabular}\\

Defining $X_{mis}$ as the missing paths of $X$ and $X_{com}=\{\mathbf{x},X_{mis}\}$, with $\mathbf{x}=\eta(\mathbf{v})$, the expectation of the complete log-likelihood $l(X_{com};\theta)$ in (\ref{llik}) can be written as:
\begin{equation}\label{is0}
\ds \mathbb{E}_{\mathbb{P}|\mathbf{x}}[l(X_{com};\theta)]=\mathbb{E}_{\mathbb{D}}
\ds \left[\frac{d\mathbb{P}}{d\mathbb{D}}(X_{mis}|\mathbf{x};\theta')l(X_{com};\theta)\right],
\end{equation}
where $\mathbb{E}_{\mathbb{P}|\mathbf{x}}$ is the expectation w.r.t. to the measure of $\ds (X_{mis}|\mathbf{x};\theta')$ under $\mathbb{P}$, and $w:=\frac{d\mathbb{P}}{d\mathbb{D}}(X_{mis}|\mathbf{x};\theta')$ is the RN derivative of that measure w.r.t. $\mathbb{D}$. The results in Lemma 4 (see Appendix B) combined with the Markov property gives that $\ds w=\prod_{i=1}^nw_{i}$, where
\begin{equation}\label{is1}
w_{i}=\kappa_{1,i}(\mathbf{x};\theta')\exp\left\{-\int_{t_{i-1}}^{t_i}\phi(s,X_{s};\theta')ds-\sum_{j=1}^{N_i}\Delta A(i,j;\theta') -\frac{(\Delta x_i-J_i)^2}{2\Delta t_i} \right\}
\prod_{j=1}^{N_i}\frac{\ell(i,j;\theta')f_g(J_{i,j};i,j,\theta')}{\lambda_i(\theta') f_i(J_{i,j};\theta')},
\end{equation}
with $\kappa_{1,i}(\mathbf{x};\theta')$ being a function that does not depend on $X_{mis}$ and $\Delta x_i=x_i-x_{i-1}$.

Now note that the complete log-likelihood $l(X_{com};\theta)$ can be written as $\sum_{i=1}^nl_i(\theta)$, where
\begin{equation}\label{ies3a}
\ds l_{i}(\theta)=A(x_{i-1};\theta)-A(x_i;\theta)-\sum_{j=1}^{N_i}\left(\Delta A(i,j;\theta)\right)-\int_{t_{i-1}}^{t_i}\phi(s,X_{s};\theta)ds +\sum_{j=1}^{N_{i}}\log\{\ell(i,j;\theta)f_{g}(J_{i,j};i,j,\theta)\}.
\end{equation}

The first natural choice of an estimator for the r.h.s. in (\ref{is0}) is
\begin{equation}\label{ies2}
\ds E_1(\theta)=\frac{1}{M}\sum_{k=1}^M\sum_{i=1}^n\left(\prod_{j=1}^nw_{j}^{(k)}\right)l_{i}^{(k)}(\theta),
\end{equation}
where $M$ is the number of Monte Carlo samples from $\ds (X_{mis}|\mathbf{x};\theta')$ under $\mathbb{D}$, and $w_{i}^{(k)}$ and $l_{i}^{(k)}$ are the values of $w_i$ and $l_i(\theta)$ for the $k$-th sampled value of $X_{mis}$.

Nevertheless, an improved estimator can be devised based on the following result.
\begin{proposition}\label{prop1}
\begin{equation}\label{ies5}
\ds \mathbb{E}_{\mathbb{P}|\mathbf{x}}\left[l(\theta)\right]= \sum_{i=1}^n\mathbb{E}_{\mathbb{D}}\left[w_{i}l_i(\theta)\right].
\end{equation}
\end{proposition}
\begin{thma}
See Appendix C.
\end{thma}
This result leads to the following Monte Carlo estimator
\begin{equation}\label{ies4}
\ds E_2(\theta)=\frac{1}{M}\sum_{k=1}^M\sum_{i=1}^nw_{i}^{(k)}l_{i}^{(k)}(\theta).
\end{equation}

Estimator $E_1$ is a global IS estimator whilst estimator $E_2$ is a local one. The latter is expected to have a much better behavior than the former. In fact, we have the following result.
\begin{proposition}\label{propvar}
\begin{equation}\label{ies6}
Var[E_2]\leq Var[E_1].
\end{equation}
\end{proposition}
\begin{thma}
See Appendix C.
\end{thma}

In order to devise an exact algorithm we need to avoid the calculation of the integrals in the expressions of $w_i$ and $l_i(\theta)$. This is achieved by firstly defining the auxiliary variables: $K_i\sim Poi(\mu_i(X;\theta')\Delta t_i)$, $U_{i,k}\sim \mathcal{U}(t_{i-1},t_i)$, $\dot{U}_i\sim\mathcal{U}(t_{i-1},t_i)$, for $k=1,\ldots,K_i$, $i=1,\ldots,n$. We also define $K=(K_1,\ldots,K_n)$, $U_i=(U_{i,1},\ldots,U_{i,K_i})$, $U=(U_1,\ldots,U_n)$ and $\dot{U}=(\dot{U}_1,\ldots,\dot{U}_n)$, and assume all the components of $K$, $U$ and $\dot{U}$ to be independent. The $\mu_i(X)$'s are functions of $X$ that must be chosen appropriately, as it is explained in the next paragraphs, to ensure the efficiency of the algorithm. Now, using the same auxiliary variable strategy from Subsection \ref{sssecmcem1} and the ideas from the Poisson estimator \citep[see][]{bpr06a}, we have the following result.
\begin{proposition}\label{propISest}
\begin{equation}\label{is7}
\ds \mathbb{E}[w_il_i(\theta)]=
\mathbb{E}\left[h_{0,i}
e^{(\mu_i(X;\theta')-b_i(X;\theta'))\Delta t_i}\mu_{i}(X;\theta')^{-K_i}\left(\prod_{k=1}^{K_i}(b_i(X;\theta')-h_{1,i,k})\right)
\left(h_{2,i}-\Delta t_i h_{3,i}\right)\right],
\end{equation}
where the expectation on the l.h.s. is w.r.t. $\ds \mathbb{D}$ and the one on the r.h.s. is w.r.t. $\ds \mathbb{D}\otimes\mathbb{A}$, with $\mathbb{A}$ being the joint probability measure of $(K,U,\dot{U})$. Also, both of those measures consider $\theta=\theta'$. The $b_i(X;\theta')$'s are functions of $X$ that must be chosen appropriately, as it is explained in the next paragraphs, to ensure the efficiency of the algorithm. Finally,
\begin{eqnarray}\label{isw2}
h_{0,i}&=&\kappa_{1,i}(X;\theta')\prod_{j=1}^{N_i}\frac{\ell(i,j;\theta')f_g(J_{i,j};i,j,\theta')}{\lambda_i(\theta')f_i(J_{i,j};\theta')} \exp\left\{-\sum_{j=1}^{N_i}\left(\Delta A(i,j;\theta')\right)-\frac{(\Delta x_i-J_i)^2}{2\Delta t_i}\right\}; \nonumber \\
h_{1,i,k}&=&\phi\left(U_{i,k},X_{U_{i,k}};\theta'\right);\;\;\;
h_{3,i}=\phi(\dot{U}_i,X_{\dot{U}_i};\theta); \nonumber \\
h_{2,i}&=&A(x_{i-1};\theta)-A(x_i;\theta)-\sum_{j=1}^{N_{i}}\left(\Delta A(i,j;\theta)\right)+\sum_{j=1}^{N_{i}}\log\left(\ell(i,j;\theta)\right)+\log\left(f_{g}(J_{i,j};i,j,\theta)\right).
\end{eqnarray}
\end{proposition}
\begin{thma}
See Appendix C.
\end{thma}

We now need to simulate $K$, $U$, $\dot{U}$, $X_{\dot{U}_i}$'s, $X_{U_{i,k}}$'s, $X_J$, $\mathbf{J}$ and ${\textbf L_{i}}$'s to obtain IS estimates of the expectations on the r.h.s. of (\ref{ies5}). Recall that the random variable ${\textbf L_{i}}$, as introduced in Subsection \ref{MCMC_1}, defines local upper and lower bounds for $X$ in $[t_{i-1},t_i]$. These are local bounds in the sense that they bound the specific path simulated from $\mathbb{D}$ and, therefore, dismisses the need for uniform (lower and upper) bounds on $\ds \phi\left(s,u;\theta'\right)$, for $u\in\mathds{R}$, $s\in[0,T]$.

Naturally, the choices of $b_i(X;\theta')$ and $\mu_i(X;\theta')$ are closely related to the efficiency of the algorithm. Based on the results from \citet{bpr06a} and \citet{prf08}, we recommend $b_i(X;\theta')=\bar{U}_i(X;\theta')$ and $\mu_i(X;\theta')=\bar{U}_i(X;\theta')-\bar{L}_i(X;\theta')$, where $\ds \bar{L}_i(X;\theta')$ and $\bar{U}_i(X;\theta')$ are, respectively, local lower and upper bounds for $\ds \phi\left(s,X_{s};\theta'\right)$, for $s\in[t_{i-1},t_i]$, which are a function of ${\textbf L_{i}}$, as it is explained in Section \ref{MCMC_1}.

We simulate the Brownian bridge component $\dot{X}_s$ of $\mathbb{D}$ by simulating standard brownian bridges and applying the linear transformation in (\ref{trans3}) to obtain the desired bridges. Defining $\dot{z}_{i,j}$ and $\dot{u}_{i,j}$ as the lower and upper bounds of $\dot{X}_s$, for $s\in(\tau_{i,j-1},\tau_{i,j})$, provided by $L_{i,j}$ as defined in Section \ref{MCMC_1}, we set $\ds\bar{H}_{i,j}=[\min\{X_{i,j-1},X_{i,j-}\}+\dot{z}_{i,j}\;,\;\max\{X_{i,j-1},X_{i,j-}\}+\dot{u}_{i,j}]$ as the respective local lower and upper bounds for $X_{s}$ in $(\tau_{i,j-1},\tau_{i,j})$. These, in turn, are used to obtain the local bounds $\bar{L}_i(X;\theta')$ and $\bar{U}_i(X;\theta')$ for $\ds \phi\left(s,X_{s};\theta'\right)$, in $[t_{i-1},t_i]$ as follows.
\begin{eqnarray}
\ds \bar{U}_i(X;\theta')&=&\sup_{j}\sup\{\phi\left(s,X_s;\theta'\right);\; s\in(\tau_{i,j-1},\tau_{i,j}),\;X_s\in \bar{H}_{i,j}\}; \label{Peb4.1}\\
\ds \bar{L}_i(X;\theta')&=&\inf_{j}\inf\{\phi\left(s,X_s;\theta'\right);\; s\in(\tau_{i,j-1},\tau_{i,j}),\;X_s\in \bar{H}_{i,j}\}. \label{Peb4.2}
\end{eqnarray}

Using the results from Proposition \ref{propISest}, we get the following estimator:
\begin{equation}\label{ies8}
\ds \dot{E}_{2}(\theta)=\frac{1}{M}\sum_{k=1}^M\sum_{i=1}^n\dot{w}_{i}^{(k)}\dot{l}_{i}^{(k)}(\theta),
\end{equation}
where $\dot{w}_{i}^{(k)}$ and $\dot{l}_{i}^{(k)}(\theta)$ are the $k$-th sample value of
\begin{eqnarray}
\ds \dot{w}_{i}&=&h_{0,i}e^{(\mu_i(X;\theta')-b_i(X;\theta'))\Delta t_i}\mu_{i}(X;\theta')^{-K_i}\prod_{k=1}^{K_i}(b_i(X;\theta')-h_{1,i,k}), \label{ies9a} \\
\dot{l}_{i}(\theta)&=&\left(h_{2,i}-\Delta t_i h_{3,i}\right). \label{ies9b}
\end{eqnarray}

The efficiency of the proposed important sampling MCEM algorithm relies heavily on the variance of the IS estimator. In particular, it is important to have a finite variance estimator, which typically relies on finite variance weights. Finite variance weights do not ensure that the estimator also has finite variance (they do whenever the target function is bounded), but weights with infinite variance will typically lead to infinite variance estimators. Besides being hard to check if the weights $\dot{w}_{i}$ have finite variance, these depend on terms $\ds \kappa_{1,i}(\mathbf{x};\theta')$ (through $f_{0,i}$) which, in turn, depend on the (unknown) transition density.
Nevertheless, we can modify $\dot{E}_2$ in a way that terms $\ds \kappa_{1,i}(\mathbf{x};\theta')$ can be ignored and also get a nice property regarding the variance of the estimator. The fact that $\mathbb{E}[\dot{w}_{i}]=1$ suggests the following estimator:
\begin{equation}\label{ies7}
\ds \dot{E}_{3}(\theta)=\sum_{i=1}^n\sum_{k=1}^M\tilde{w}_{i}^{(k)}\dot{l}_{i}^{(k)}(\theta),
\end{equation}
where $\ds {\tilde{w}_{i}}^{(k)}=\frac{\dot{w}_{i}^{(k)}}{\sum_{k=1}^M\dot{w}_{i}^{(k)}}$.
Estimator $\dot{E}_3$ is biased but has some nice properties as stated in the following lemma.

\begin{lemma}\label{theofveme}
$\dot{E}_3$ is a strongly consistent estimator for the expectation from the E-step and has finite variance whenever $\dot{l}_i(\theta)$ has finite variance, for all $i$, under $\ds \mathbb{D}\otimes\mathbb{A}$.
\end{lemma}
\begin{thma}
See Appendix C.
\end{thma}
Although $\dot{E}_3$ is biased, Lemma \ref{theofveme} implies that it is asymptotically unbiased. Furthermore, the number of Monte Carlo samples $M$ is expected to be typically large enough to make the bias negligible.

In the general case in which the diffusion coefficient depends on unknown parameters, we consider the complete log-likelihood $l_i(\theta)$ to be as in (\ref{comllik}) and obtain the transformed bridges $\dot{X}$ by applying the transformation in (\ref{trans2}).

Finally, we also construct an estimator for the covariance matrix of the MLE, which is presented in Appendix D.

\subsection{An infinite-dimensional Barker's MCMC}\label{idMCMCsec}

If at least one of the assumption of the JBEA algorithm (see Appendix A) is not satisfied, we are unable to use it to exactly simulate bridges from the jump-diffusion $X$ and, as a consequence, we are unable to find a finite-dimensional representation of those bridges for which we know and can simulate from the full conditional distribution of the parameters given this representation.
In this case, we can devise a Bayesian solution based on the (infinite-dimensional) complete bridges $V_{mis}$ (as define in Subsection \ref{MCEMudc}). We consider the general case in which the diffusion coefficient $\sigma(V_{s-};\theta)$ depends on unknown parameters and propose a MCMC algorithm that iterates between the update of the two following blocks via accept/reject algorithms:
\begin{equation}\label{mcmcfc2}
\ds \left(V_{mis}|\theta,\textbf{v}\right)\;\;\mbox{and}\;\;\left(\theta|V_{mis},\textbf{v}\right).
\end{equation}
The first block cannot be sampled exactly from its full conditional distribution due to the intractability of JBEA. In order to perform inference without resorting to discrete approximations we need to sample both $V_{mis}$ and $\theta$ in a Markov chain with the required invariant distribution, by unveiling only a finite-dimensional part of $V_{mis}$. That can be achieved by constructing a MCMC algorithm with accept/reject steps for both blocks - $V_{mis}$ and $\theta$, and adopting a Barker's accept/reject algorithm, instead of the traditional Metropolis-Hastings one. The particular form of the Barker's acceptance probability allows us to devise a Bernoulli factory algorithm to sample events with the respective (intractable) acceptance probabilities by unveiling only a (random) finite-dimensional representation of $V_{mis}$.

The main contribution of this algorithm when compared to the one from Subsection \ref{MCMC_1} is that it can be applied to a quite general class of jump-diffusion processes, including processes with unbounded drift and/or unbounded jump rate and unbounded function $(\alpha^2+\alpha')$. 

\subsubsection{Sampling the missing paths}

The missing paths of $X$ are sampled via a Metropolis type algorithm but using the Barker's acceptance probability instead of the traditional Metropolis-Hastings one.
Barker's algorithm was proposed in \citet{Barker} and, like the MH algorithm, defines a Markov chain with a given target invariant distribution. On each iteration of the chain, a proposal value is simulated from an arbitrary distribution and accepted according to an acceptance probability that preserves detailed balance. The probability to go from $x$ to $y$ when proposing from $q(y|x)$ is given by
\begin{equation}\label{apbak}
\ds a_{B}(x,y)=\frac{\pi^*(y)q(x|y)}{\pi^*(x)q(y|x)+\pi^*(y)q(x|y)},
\end{equation}
where $\pi^*$ is the invariant distribution of the chain. Barker's method is not as popular as the prolific Metropolis-Hastings algorithm since it is easy to demonstrate that Barker is uniformly dominated by MH in terms of its convergence properties \citep[see][]{Peskun}. However  it is also easy to see that Barker is no worse than twice as slow as Metropolis-Hastings and its convergence properties are broadly comparable \citep{LRB}. More specifically, we can show that $a_{MH}(x,y)/2 \leq a_{B}(x,y) \leq a_{MH}(x,y)$, for all $(x,y)$ and, as it is shown by \citet{LRB}, for all measurable function $f\in L^2(\pi^*)$ for which a square root central limit theorem (CLT) holds for the MH chain with CLT asymptotic variance $\sigma_{MH}^2$, a corresponding CLT holds for $f$ and the Barker's chain with CLT asymptotic variance $\sigma_{B}^2$, such that $\sigma_{MH}^2\leq \sigma_{B}^2 \leq 2\sigma_{MH}^2+\sigma_{\pi^*}^2$, where $\sigma_{\pi^*}^2:=Var_{\pi^*}(f)$.
For us, the smoothness of the Barker's acceptance probability function turns out to be crucial in obtaining a feasible algorithm.

The block $V_{mis}$ is broken into $n$ blocks which consists of $V_{mis}$ restricted to each of the $n$ time intervals $(t_{i-1},t_i)$. Due to the Markov property, the full conditional distribution of each of these blocks depend only on $\theta$ and $(v_{i-1},v_{i})$, which means that these $n$ sampling steps on each iteration of the MCMC algorithm are conditionally independent and can be performed in parallel.
The Barker's proposal for $V_{mis}$ in $(t_{i-1},t_i)$ is sampled from $\mathbb{D}$, as defined in Section \ref{ssecismcem}, using Algorithm 1 from that same section. Note that the initial and end values that the bridges - $x_{i-1}(\theta)$ and $x_i(\theta)$, are functions of $\theta$.

For a given interval $(t_{i-1},t_i)$, let $X^{(k-1)}$ be the current state of $V_{mis}$ in the MCMC chain and a new proposal jump-diffusion bridge $X^{(k)}$ is drawn from $\mathbb{D}$. It is convenient to use the proposal measure also as a dominating measure to obtain the acceptance probability which, from Lemma 4 (see Appendix B), is given by
\begin{equation}\label{apbak2}
\ds \alpha_X = \ds\frac{\ds\frac{d\mathbb{P}}{d\mathbb{D}}(X^{(k)}|\mathbf{x}(\theta))}{\ds\frac{d\mathbb{P}}{d\mathbb{D}}(X^{(k-1)}|\mathbf{x}(\theta))+\ds\frac{d\mathbb{P}}{d\mathbb{D}}(X^{(k)}|\mathbf{x}(\theta))}= \frac{s_k(X^{(k)})p_k(X^{(k)})}{s_{k-1}X^{(k-1)}p_{k-1}X^{(k-1)}+s_k(X^{(k)})p_k(X^{(k)})},
\end{equation}
where
\begin{equation}\label{apbak4}
\ds s_{k}(X^{(k)})=\exp\left\{-\sum_{j=1}^{N_i}\left(\Delta A(i,j;\theta)\right)-\mathcal{I}_{a_i}(\theta)-\frac{(\Delta x_i(\theta)-J_i)^2}{2\Delta t_i}\right\} \prod_{j=1}^{N_i}\frac{\ell(i,j;\theta)f_g(J_{i,j};i,j,\theta)}{\lambda_i(\theta)f_i(J_{i,j};\theta)},
\end{equation}
\begin{equation}\label{apbak5}
\ds p_{k}(X^{(k)})=\exp\left\{-\int_{t_{i-1}}^{t_i}\phi(s,X_{s};\theta)-a_i(s;\theta)ds\right\},
\end{equation}
for $\Delta x_i(\theta)=x_i(\theta)-x_{i-1}(\theta)$, $\ds \mathcal{I}_{a_i}=\int_{t_{i-1}}^{t_i}a_i(s;\theta)ds$ and $a_i(s;\theta)$ being any (local) lower bound satisfying
\begin{equation}\label{apbak6}
\ds a_i(s;\theta)\leq\phi(s,X_{s};\theta),\;\mbox{for }s\in(t_{i-1},t_i).
\end{equation}
The expressions for $s_{k-1}(X^{(k-1)})$ and $p_{k-1}(X^{(k-1)})$ are obtained by replacing $X^{(k)}$ by $X^{(k-1)}$ in (\ref{apbak4})-(\ref{apbak6}).

The choice of $a_i(s;\theta)$ has direct impact on the efficiency of the algorithm, as it is discussed further ahead in Subsection \ref{secsampar}. If the function $\phi$ is not uniformly bounded below, $a_i(s;\theta)$ can only be a local bound. We can set, for example, $\ds a_i(s;\theta)=\bar{L}_i(X;\theta')$, for $s\in(t_{i-1},t_i)$, as defined in (\ref{Peb4.2}). This choice ought to work well in most cases. A more refined bound, however, may be obtained by adopting a time-dependent bound $a_i(s;\theta)$ (see Appendix F).

To perform the accept/reject step of the algorithm, we need to simulate a $Bernoulli(\alpha_X)$ random variable. Most importantly,
we have to use only a finite-dimensional representation of the missing paths to do it. That is achieved by the \emph{Two-Coin algorithm}
given in the following proposition.

\begin{proposition}
Suppose we want to simulate a $Bernoulli(\alpha_X)$ random variable, where $s_{k}$ and $s_{k-1}$ are known positive numbers
and it is possible to simulate events of unknown probabilities given by $p_{k}$ and $p_{k-1}$.
The following algorithm outputs an exact draw of this Bernoulli random variable:\\
\\
\begin{tabular}[!]{|l|}
\hline\\
\parbox[!]{13cm}{
\texttt{
{\bf Two-Coin algorithm} {\scriptsize
\begin{enumerate}\label{2ca}
\setlength\itemsep{-0.5em}
  \item Sample $C_1$ from $\{0,1\}$, where $\ds P(C_1=1)=\frac{s_{k}}{s_{k-1}+s_{k}}$;
  \item if $C_1=1$, sample $C_2$ from $\{0,1\}$, where $\ds P(C_2=1)=p_{k}$;
  \begin{itemize}
  \setlength\itemsep{-0.5em}
    \item if $C_2=1$, output 1;
    \item if $C_2=0$, go back to 1;
  \end{itemize}
  \item if $C_1=0$, sample $C_2$ from $\{0,1\}$, where $\ds P(C_2=1)=p_{k-1}$;
  \begin{itemize}
  \setlength\itemsep{-0.5em}
    \item if $C_2=1$, output 0;
    \item if $C_2=0$, go back to 1.
    \end{itemize}
\end{enumerate}}
}}\\  \hline
\end{tabular}\\
\end{proposition}
\begin{thma}
Let $q$ be the probability that there is no output in one trial of $C_1$ and $C_2$, that is $q=\frac{s_k(1-p_k)+s_{k-1}(1-p_{k-1})}{s_k+s_{k-1}}$.
Then, the probability that the algorithm outputs 1 is $\frac{s_k}{s_k+s_{k-1}}p_k\sum_{i=0}^{\infty}q^i=\frac{s_kp_k}{s_kp_k+s_{k-1}p_{k-1}}$.
The fact that the algorithm is a Geometric experiment implies that it finishes after a finite number of loops with probability 1, which concludes the proof.
\end{thma}

In order to simulate events of probability $p_k(X^{(k)})$ and $p_{k-1}(X^{(k-1)})$ we use the Poisson Coin algorithm described in Appendix A, for which we require an upper bound $r_i(\theta)$ for the integrand $\ds \phi(s,X_{s-};\theta)-a_i(s;\theta)$ in (\ref{apbak5}) in the interval $[t_{i-1},t_i]$. This can be a local bound, which explains why a uniform upper bound on that function is not required here. Setting, for example, $\ds a_i(s;\theta)=\bar{L}_i(X;\theta')$, for $s\in(t_{i-1},t_i)$, and $\ds r_i(\theta)=\bar{U}_i(X;\theta')-\bar{L}_i(X;\theta')$, as defined in (\ref{Peb4.1}) and (\ref{Peb4.2}), should work well in most cases. More refined bounds, however, can be obtained by adopting a time-dependent bound $a_i(s;\theta)$ (see Appendix F).

It is important to notice that all the values of the proposed path unveiled along the Two-Coin algorithm need to be kept, meaning that the simulation of the path in any new time instant has to be conditioned on all the already unveiled points.

Let $X_{\Psi}$ be the process $X$ at a finite random collection of time instants which, for each interval $[t_{i-1},t_i]$, are sampled in the Poisson Coin algorithm used in all the loops of the Two-Coin algorithm. Each Barker's step for $V_{mis}$ stores a finite-dimensional representation of the missing path, consisting of $\mathbf{J}$, $X_{J}$, ${\textbf L_{i}}$ and $X_{\Psi}$ in each interval $[t_{i-1},t_i]$, $i=1,\ldots,n$. It is crucial though to be able to simulate further points, given this skeleton, on the $\theta$ step of the MCMC algorithm, as we discuss in Subsection \ref{secsampar} below.

\subsubsection{Sampling the parameters}\label{secsampar}

Typically, the full conditional distribution of the parameters will depend on the unknown integral in (\ref{fullcpar5}). In order to minimise the complexity of the algorithm, the parameters should firstly be separated into two blocks:
the first - $\theta_1$,  consisting of those parameters whose full conditional distributions depend on the integral in (\ref{fullcpar5}), and the second block - $\theta_2$, consisting of the remaining parameters. Parameters in $\theta_2$ are sampled as in an ordinary tractable MCMC - they may be broken into smaller blocks, sampled directly from the full conditional or via MH steps.
Parameters in $\theta_1$ are sampled via Barker's step, which may be performed separately for sub-blocks. The full conditional density of $\theta$ and, consequently, of any subvector of its coordinates is given by
\begin{equation}\label{fullcpar1}
\ds \pi(\theta|V_{mis},\textbf{v})\propto\pi(V_{mis},\textbf{v}|\theta)\pi(\theta),
\end{equation}
where $\ds \pi(V_{mis},\textbf{v}|\theta)$ is given in Lemma \ref{comlliklem}.

Proposals for $\theta_1$ are drawn from a symmetric random walk, $\theta_{1}^{(k)}=\theta_{1}^{(k-1)}+\epsilon$, where $\epsilon$ is a r.v. symmetric around 0 with a covariance matrix $\Sigma$ properly tuned to obtain suitable acceptance rates (varying from around 0.27, for unidimensional $\theta_1$, to around 0.16 for dimension 5+ \citep{DV_barker}). A move $\theta_{1}^{(k-1)}\rightarrow\theta_{1}^{(k)}$ is accepted with probability:
\begin{equation}\label{fullcpar2}
\ds \alpha_{\theta}=\frac{\pi(\theta^{(k)}|V_{mis},\textbf{v})}{\pi(\theta^{(k-1)}|V_{mis},\textbf{v})+\pi(\theta^{(k)}|V_{mis},\textbf{v})}=
\frac{s_k(\theta^{(k)})p_k(\theta^{(k)})}{s_{k-1}(\theta^{(k-1)})p_{k-1}(\theta^{(k-1)})+s_k(\theta^{(k)})p_k(\theta^{(k)})},
\end{equation}
where
\begin{eqnarray}\label{fullcpar4}
\ds s_k(\theta^{(k)})&=&\pi(\theta^{(k)})\exp\left\{A(x_n(\theta^{(k)});\theta^{(k)})-A(x_0(\theta^{(k)});\theta^{(k)})
-\sum_{i=1}^n\mathcal{I}_{a_i}(\theta^{(k)}) \right\} \nonumber \\
&\times& \prod_{i=1}^{n}\left[\exp\left\{-\sum_{j=1}^{N_i}\Delta A(i,j;\theta^{(k)})\right\} \prod_{j=1}^{N_i}\ell(i,j;\theta^{(k)})f_{g}(J_{i,j};i,j,\theta^{(k)})\right] \nonumber \\
&\times&\prod_{i=1}^{n}\left[\sigma(V_{t_i};\theta^{(k)})^{-1}f_{N}(X_{i,1-};x_{i-1}(\theta^{(k)}),\tau_{i,1}-t_{i-1}) f_{N}(x_i(\theta^{(k)});X_{i,N_i},t_i-\tau_{i,{N_i}})\right],
\end{eqnarray}
\begin{equation}\label{fullcpar5}
\ds p_k(\theta^{(k)})=\exp\left\{-\ds\int_{0}^T\dot{\phi}(s,\dot{X}_{s};\theta^{(k)})-\dot{a}(s;\theta^{(k)})ds\right\},
\end{equation}
\begin{equation}\label{fullcpar6a}
\ds \dot{a}(s;\theta^{(k)})=\dot{a}_{i,j}(s;\theta^{(k)}),\;\mbox{for }s\in[\tau_{j,i-1},\tau_{j,i}],
\end{equation}
\begin{equation}\label{fullcpar7}
\ds \dot{a}_{i,j}(s;\theta)\leq\dot{\phi}(s,\dot{X}_{s};\theta),\;\mbox{for }s\in[\tau_{j,i-1},\tau_{j,i}],
\end{equation}
where $\theta^{(k)}$ has values $\theta_{1}^{(k)}$ for $\theta_1$ and the current value of the chain for $\theta_2$. The expressions for $s_{k-1}(\theta^{(k-1)})$ and $p_{k-1}(\theta^{(k-1)})$ are obtained by replacing $\theta^{(k)}$ by $\theta^{(k-1)}$ in (\ref{fullcpar4})-(\ref{fullcpar6a}).

The efficiency of this sampling step depends on the efficiency of the Two-Coin algorithm which ultimately relies on the probabilities of success of $C_2$, $p_{k-1}$ and $p_k$.
The smaller these probabilities are, the higher is the expected number of trials per iteration. Moreover, at every trial, the missing paths have to be unveiled at extra time points, which also increases the computational cost.
The optimisation of $p_{k-1}$ and $p_k$ is related to the optimisation of the lower bounds $\dot{a}_{i,j}$ from (\ref{fullcpar7}).

We simulate the second coin piecewise by simulating a sequence of ``sub-coins", forward in time from 0 to $T$, with probability
\begin{equation}\label{fullcpar5i}
\ds p_{k,i,j}=\exp\left\{-\ds\int_{\tau_{i,j-1}}^{\tau_{i,j}}\dot{\phi}(s,\dot{X}_{s};\theta^{(k)})-\dot{a}_{i,j}(s;\theta^{(k)})ds\right\}
\end{equation}
or $p_{k-1,i,j}$, which replaces $\theta^{(k)}$ by $\theta^{(k-1)}$ in (\ref{fullcpar5i}). If we get 0 at some point, the remaining sub-coins do not need to be simulated and the Two-Coin algorithm starts a new loop.

In order to simulate each sub-coin $p_{k,i,j}$, we apply the Poisson Coin algorithm by simulating a Poisson process with rate $\dot{r}_{i,j}(\theta^{(k)})$, where $\dot{r}_{i,j}(\theta)$ is a local upper bound for the integrand $\ds \dot{\phi}(s,X_{s-};\theta)-\dot{a}(s;\theta)$ in (\ref{fullcpar5}) in the interval $[\tau_{i,j-1},\tau_{i,j}]$. An efficient solution to obtain $\dot{r}_{i,j}(\theta)$ is presented in Appendix F.

The overall MCMC algorithm is the following.\\
\\
\begin{tabular}[!]{|l|}
\hline\\
\parbox[!]{13cm}{
\texttt{\scriptsize
{\bf Barker's MCMC for jump-diffusions}
\begin{enumerate}\label{2ca}
\setlength\itemsep{-0.5em}
  \item Provide initial values for all the parameters $\theta$;
  \item make k=1;
  \item sample $V_{mis}^{(k)}$ (retrospectively and including ${\textbf L_{i}}$) once on every interval between consecutive observations via Barker's by proposing from the measure $\mathbb{D}$ and accepting with probability $\alpha_X$ in (\ref{apbak2}) using the Two-Coin algorithm;
  \item sample each block of $\theta_{1}^{(k)}$ via Barker's by proposing from a symmetric random walk and accepting with probability $\alpha_{\theta}$ in (\ref{fullcpar2}) using the Two-Coin algorithm;
  \item sample the remaining parameters $\theta_{2}^{(k)}$ directly from their full conditionals or via Gaussian random walk MH;
  \item to continue running the chain, make $k=k+1$ and GOTO 3, otherwise, STOP.
\end{enumerate}
}}\\  \hline
\end{tabular}\\

\subsubsection{Improving the MCMC algorithm}\label{impbkmcmc}

The efficiency of the proposed MCMC algorithm in terms of convergence and computing cost depends on many factors.  Here we shall discuss strategies for improvements.

In order to avoid numerical problems when computing the probability of $C_1$ in the Two-Coin algorithm, compute $\ds (1+\exp(\log(s_{k-1})-\log(s_k)))^{-1}$.

A way to improve the mixing of the chain is to add an extra step to the Gibbs sampler to sample the missing (continuous) paths between observation and jump times using the EA algorithm \citep[see][]{bpr06b}.
We only require the extra condition that $(\alpha^2+\alpha')(u)$ is bounded below for all $u$ in the state space of $X$. EA performs rejection sampling by proposing from a Brownian bridge with the same initial and ending values as the target diffusion bridge and accepting with probability given by
$\ds \exp\left\{-\int_{t_{i-1}}^{t_i}\left(\frac{\alpha^2+\alpha'}{2}\right)(X_{s})-l\;ds\right\}$, where $l=\inf_u\left(\frac{\alpha^2+\alpha'}{2}\right)(u)$.
Since this algorithm will output an exact draw of the missing paths between observation and jump times, it guarantees that these bridges are updated on every iteration of the chain at least once. This strategy ought to improve the mixing of the chain considerably. Furthermore, it eliminates the problem of accumulating too many bridge points along the iterations of the chain due to rejections on the Barker's step. It is recommended to add one EA step after each Barker's step - for $X$ or $\theta$-blocks. The example in Subsection \ref{sim2subsec} implements this extra update step.

Further improvements in the chain mixing could be obtained by modifying
the Barker's step for $V_{mis}$, which is the only step where the jump process is updated. One idea is to perform multiple steps of this type.
Another simple and virtually costless strategy is to adopt a pilot analysis to tune the jump rate based on the average number of jumps in each interval. Moreover, if one has reasonable choices for initial values of the parameters it may be a good idea to warm up the chain on the first iterations before start updating the parameters.

In order to increase the success probability of $C_2$ in the Two-Coin algorithm we need to improve the lower bound $\dot{a}(s,\theta)$, which can be done by tightening the lower and upper bounds for the missing paths in $V_{mis}$. That is achievable by performing what we call the \emph{layer refinement algorithm} which, instead of simulating upper and lower bounds (through ${\textbf L_{i}}$) for the proposal standard bridges between the merged observation and jump times, simulates upper and lower bounds for shorter intervals. For a constant $m\in\mathds{N}$ and a time interval $(\tau_{i,j-1},\tau_{i,j})$, we first simulate a standard BB at times $\tau_{i,j-1}+k(\tau_{i,j}-\tau_{i,j-1})/m$, for $k=1,\ldots,m-1$, and then obtain upper and lower bounds for each of the sub-intervals. This strategy will provide tighter bounds for the standard bridge and, consequently, for the $X$ path. In particular, for a standard bridge of length $t$, the range is $\mathcal{O}(\sqrt{t})$. One may also set different refinement levels along the intervals of consecutive observations. It is typically straightforward to identify the relation between the cost of the Two-coin Algorithm and the values assumed by the process $X$, so the level of refinement may be set as a function of the extreme observations of each interval. The detailed algorithms to simulate ${\textbf L_{i}}$ and obtain the bounds for the $X$ path are presented in Appendices E and F. A cheaper version of the EA step described above consists of updating the missing continuous bridges between observation, jump and refinement points, conditioned on the auxiliary variable ${\textbf L_{i}}$.

The layer refinement algorithm suggests that the algorithm that simulates only the first coin $C_1$ may be seen as a discretised method and it is the second coin that guarantees the exactness of the algorithm. In that case, the larger $m$ is the smaller the error due to the discretisation.

Two other strategies may increase the probability of the second coins. The first one is to break $\theta_1$ into smaller blocks (the extreme case being one parameter per block). This may help in the sense that a smaller block may simplify the function inside the integral in (\ref{fullcpar5}) and allow for a more efficient lower bound $a(s;\cdot)$.
The second strategy is to divide the numerator and denominator of the Barker's acceptance probability by $p_{k-1}$ or $p_k$. This will make the probability of one of the second coins equal to 1 and possibly increase the probability of the other one. In order to choose between $p_{k-1}$ and $p_k$ one has to look at the resulting ratio of $p$'s and recognise which choice has a ratio of the form $s_0\int_{0}^Th(X_s)ds$, where $s_0$ is computable and $h$ is a non-negative function. This choice may vary between iterations of the chain, depending on the proposed value of the parameters. If this strategy is adopted, one of the $C_2$ coins will have success probability 1 and the other one will typically get smaller as the size of the random walk step increases. For this reason, it may be wise to truncate the random walk proposal for $\theta_1$ (if it is Gaussian) between say $\pm3.5$ or $\pm4$ standard deviations to avoid the algorithm from collapsing after an average number of $1/\epsilon$ iterations - $\epsilon$ being the probability of proposing extreme values. This would have very little effect on the algorithms's convergence properties. Two other reasonable strategies are to use a uniform random walk proposals and to perform multiple updates using smaller variance proposals. The multiple proposals ought to compensate the slower mixing but with considerable gains in computational cost.

\subsection{Practical implementation}\label{mipi}

Identifiability is a particular problem when dealing with jump-diffusions. It is crucial to have enough information to distinguish well between continuous and jump variation. Practical strategies to tackle the problem include fixing some of the parameters at reasonable values, which is not always easy, or using informative priors under a Bayesian approach. A general idea that should always be considered is that of admitting the least possible number of jumps necessary to get a good fit so that the jumps only occur when a pure diffusion process is not flexible enough to model the phenomenon of interest. This not only mitigates the identifiability problem but also favors the interpretability of jumps and parameters and, finally, ought to improve computational efficiency.

Computational cost is another important issue when dealing with the algorithms proposed in this paper. Since the MCEM and MCMC algorithms from Section \ref{JBEAalgs} are considerably cheaper than the algorithms presented in Section \ref{inf2}, we may consider some practical strategies to allow the use of the former ones. For example, we may occasionally truncate the jump rate. However, this strategy ought to be adopted with care as it may seriously compromise the analysis. On one hand, a severe truncation could significantly compromise desirable properties of the original (without truncation) model. On the other hand, a conservative bound (that uses a truncation value that is expected to be hardly reached by the model) is bound to seriously compromise the computational cost of the algorithm. The bounds for the jump rate and drift obtained from the truncation are used to specify $\lambda_0$ and $\kappa_{0}$, as defined in Appendix A. The use of conservative bounds then will lead to low acceptance probabilities for the JBEA algorithm.

We can also reduce the computational cost of JBEA by making its proposal as similar as possible to the target. This is also a good idea when using the algorithms from Section \ref{inf2} - it would reduce the variance of the weights in the ISMCEM algorithm and increase the success probability of $C_2$ in the MCMC one. The idea is basically to make the proposal jump process (jump rate and jump size distribution) depend on time and/or on the initial and ending values of the interval whenever the target jump process is time and/or state dependent. The time dependence can always be mimicked from the target and the state replaced by a function of the initial and ending values, for example, their mean.


\section{Simulated examples}

In this section we present results from some simulated examples. Firstly, we present an example with bounded drift and bounded jump-rate to apply the algorithms from Section \ref{JBEAalgs}. Secondly, the algorithms from Section \ref{inf2} are applied to an unbounded drift example. Three data sets are simulated from each model and analysed with the respective algorithms. The three replications in each example present similar results when applying the proposed methodologies. For that reason, results for only one of them are reported here whilst results for the other ones are reported in Appendix H.

\subsection{An introductory example}\label{ss51}

We consider the following model:
\begin{eqnarray}\label{mdsim1}
\ds dV_s &=& -\tanh(V_s-\delta)ds +\sigma dW_s+dJ_s,\;\;\;\;V_0=v,\;\;s\in[0,1000]  \\
  \lambda_1(s;V_{s-})&=&\lambda(1-\tanh^2(V_s-\delta)),\;\;\;\;\; g_1(Z_j,V_{t_j-}) =  Z_j \sim \mathcal{N}(\mu,\tau^2), \nonumber
\end{eqnarray}
with $(\delta,\sigma^2,\lambda,\mu,\tau^2)=(0,\;1,\;0.1,\;2,\;0.35^2)$. We chose these parameters to be potentially problematical for the algorithm, as there will be identifiability problems in distinguishing the presence of jumps from continuous volatility. The MCEM algorithm
of Subsection \ref{MCEMudc} and the MCMC algorithm of Subsection \ref{MCMC_1} are applied to the same dataset and results are presented in Tables \ref{tabsim1} and \ref{tabsim2}, respectively. The M-step from the MCEM algorithm is performed numerically via the quasi-Newton method BFGS \citep[see, for example,][Section 3.2]{BFGS}.

The data set presents 56 jumps, with mean 2.125 and variance 0.104. The distance between the MLE estimates and the posterior means in terms of the respective posterior standard deviations (how many posterior s.d.'s the MLE estimate is from the posterior mean) are: 0.51 for $\delta$, 0.11 for $\sigma^2$, 0.73 for $\lambda$, 0.83 for $\mu$ and 1.06 for $\tau^2$.

\begin{table}[!h]
\caption{\label{tabsim1}Iterations from the MCEM algorithm. 1000 Monte Carlo samples are used in all iterations. Results obtained with other initial conditions (away from the real values) suggest the presence of local modes.}
\centering
\fbox{\scriptsize
\begin{tabular}{c|ccccc}
  Iteration        & $\delta$ & $\sigma^2$ & $\lambda$ & $\mu$ & $\tau^2$ \\ \hline
  0                & 0.5    & 1.5    & 0.2  & 2  & 0.25 \\
  1                & 0.149  & 1.299  & 0.104  & 1.771 & 0.212  \\
  5                & 0.078  & 1.128  & 0.096  & 1.685 & 0.197  \\
  20               & 0.023  & 1.031  & 0.128  & 1.770 & 0.203  \\
  100              & 0.005  & 0.996  & 0.133  & 1.848 & 0.175 \\
  200              & 0.008  & 0.996  & 0.128  & 1.889 & 0.146 \\
  500              & 0.019  & 1.001  & 0.120  & 1.942 & 0.100 \\ \hline
 real              & 0  & 1  & 0.1  & 2  & 0.1225
\end{tabular}}
\end{table}

\begin{table}[!h]
\caption{\label{tabsim2}Posterior statistics from the MCMC output. Uniform improper priors are adopted for all parameters. The prior of $\tau^2$ had to be truncated to be above 0.008 to avoid getting trapped in small values. The chain runs for 500$k$ iterations. The parameters are jointly sampled using an adaptive Gaussian random walk MH step which had a 0.31 acceptance rate. Trace plots suggest convergence has been achieved. Parameters $\mu$ and $\tau^2$ present two marginal modes each.}
\centering
\fbox{\scriptsize
\begin{tabular}{c|ccccc}
              & $\delta$ & $\sigma^2$ & $\lambda$ & $\mu$ & $\tau^2$ \\ \hline
  Mean        & -0.025  & 0.993  & 0.179  & 1.69  & 0.291 \\
  Median      & -0.021  & 0.991  & 0.164  & 1.67  & 0.275 \\
  Mode        & -0.023  & 0.987  & 0.135  & 1.54 / 1.66  & 0.025 / 0.280 \\
  St. Dev.    &  0.086  & 0.068  & 0.080  & 0.30  & 0.180 \\ \hline
  real        &  0      & 1      & 0.1    & 2     & 0.1225
\end{tabular}}
\end{table}

\begin{figure}[!h]
\centering
   \includegraphics[scale=0.3]{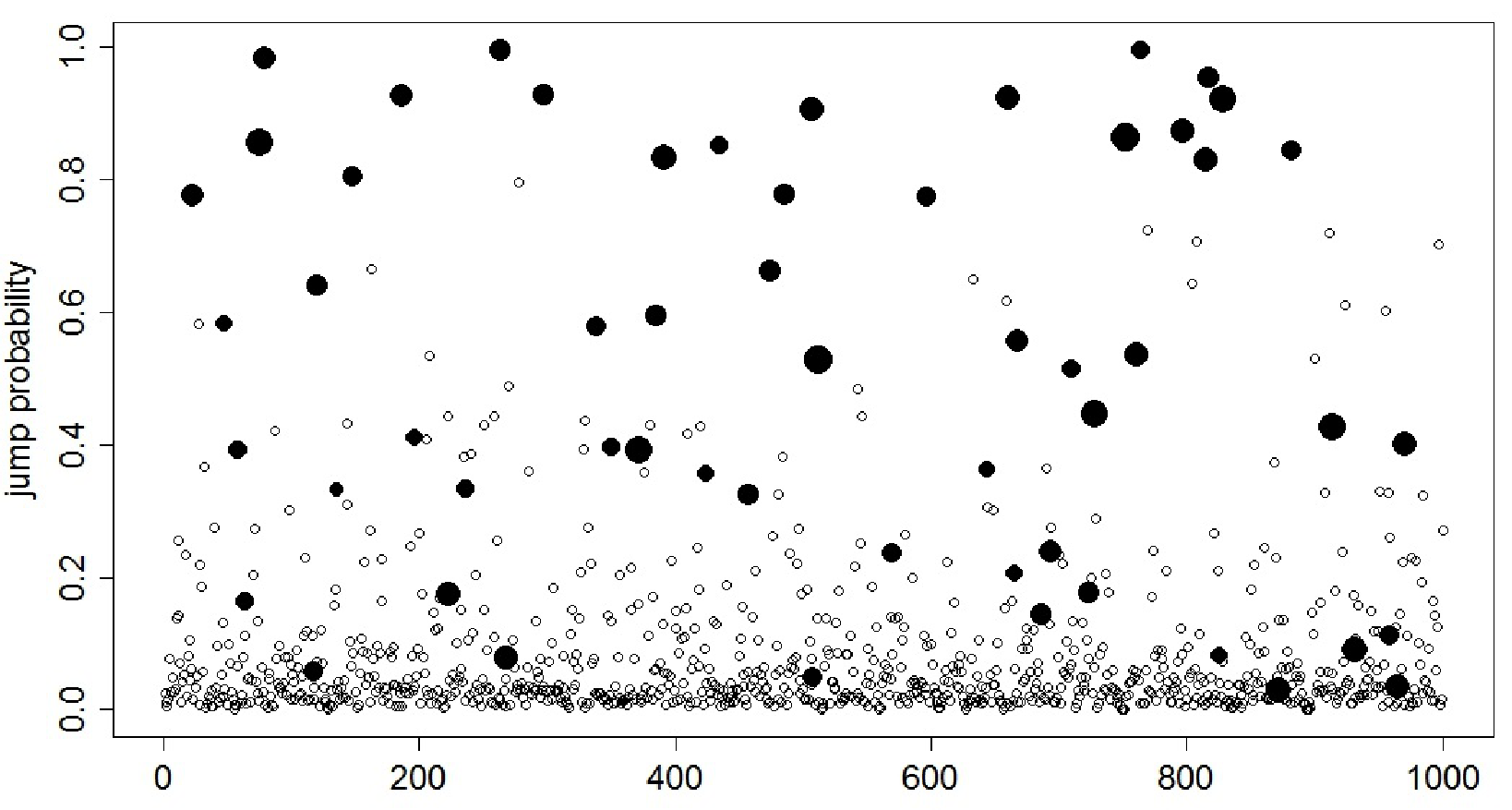}
\caption{Posterior probability of jump between each pair of consecutive observations. Solid circles represent intervals where a jump really exists and have size proportional to the size of the jump.}\label{figsim2}
\end{figure}

The MCMC algorithm also outputs a sample from the posterior distribution of the jump process. Some interesting posterior statistics can be obtained from this distribution. Figure \ref{figsim2} shows an example.

We also run the algorithms fixing parameters $\mu$ and $\tau^2$ at their real values. Table \ref{tabsim3} shows the results obtained for the two algorithms. They reinforce the issues concerning identifiability - note the improvement in the estimation of the jump rate by the MCMC algorithm. Without fixing those parameters, the posterior distribution estimates a higher average number of jumps with smaller average size.

\begin{table}[!h]
\caption{\label{tabsim3}Results for the case where $\mu$ and $\tau^2$ are fixed at their real values. The specifications of the two algorithms are the same as in the previous run. The output of the MCEM algorithm corresponds to iteration 55, for which it has clearly converged.}
\centering
\fbox{\scriptsize
\begin{tabular}{c|ccc}
              & $\delta$ & $\sigma^2$ & $\lambda$  \\ \hline
  MCEM        &  0.024   & 1.004  & 0.113   \\ \hline
  MCMC & & & \\
  Mean        &  0.021  & 1.009  & 0.117   \\
  Median      &  0.022  & 1.007  & 0.115   \\
  Mode        &  0.024  & 1.004  & 0.113   \\
  St. Dev.    &  0.064  & 0.062  & 0.028   \\
  95\% Cred. Int. & (-0.107,0.146) & (0.895,1.137) & (0.066,0.177) \\ \hline
  real        &  0      & 1      & 0.1
\end{tabular}}
\end{table}

\subsection{An unbounded drift example: the Ornstein Uhlenbeck process}\label{sim2subsec}

We now present a simulated example where the drift is unbounded and the methodology from Section \ref{inf2} is applied. We consider the following model:
\begin{eqnarray}\label{mdsim2}
\ds dV_s &=& -\rho(V_s-\mu)ds + dW_s+dJ_s,\;\;\;\;V_0=v,\;\;s\in[0,500]  \\
  \lambda_1(s;V_{s-})&=&\lambda,\;\;\;\;\; g_1(Z_j,V_{t_j-}) =  Z_j \sim Exp(\theta), \nonumber
\end{eqnarray}
with $(\rho,\mu,\lambda,\theta)=(1,\;0,\;0.07,\;1)$. Identifiability problems are likely to occur due to the ambiguity involving variation of the continuous part and small jumps from the Exponential distribution. We consider the (constant) diffusion coefficient to be known. Although the proposed methodologies can be used to estimated parameters in the diffusion coefficient, identifiability issues would be severely worsen. In real application, in which assuming the diffusion coefficient to be known may be not reasonable, we can use some specific strategies to mitigate the identifiability problem as it is done for the example in Section \ref{appsect}.

The data set presents 36 jumps, with mean (1/0.733)=1.364. Results from the MCEM algorithm of Subsection \ref{ssecismcem} are presented in Table \ref{tabsim4}, considering the estimator $\dot{E}_3$. The M-step is performed analytically. An algorithm based on estimator $E_1$ for fixed $\lambda$ was also implemented and the Monte Carlo variance was too high even for $8\times10^5$ samples.

\begin{table}[!h]
\caption{\label{tabsim4}Iterations from the MCEM algorithm using estimator $\dot{E}_3$.}
\centering
\fbox{\scriptsize
\begin{tabular}{c|ccccc}

  Iteration        & $\rho$ & $\mu$ & $\lambda$ & $\theta$ & MC samples ($\times10^3$)  \\ \hline
  0                &   1.5  &  -0.5  &    0.5    &    1.5   &   \\
  1                & 1.459  & -0.231 &  0.651    &  1.245   &   20  \\
  10               & 1.185  & -0.105 &  0.605    &  1.667   &   50  \\
  100              & 1.053  &  0.021 &  0.281    &  1.486   &   50  \\
  200              & 1.005  &  0.099 &  0.094    &  0.926   &   50  \\
  300              & 0.989  &  0.117 &  0.066    &  0.801   &   50  \\
  312              & 0.996  &  0.114 &  0.067    &  0.748   &   600 \\
  316              & 1.001  &  0.114 &  0.066    &  0.760   &   600 \\
  \hline
 real              & 1      & 0      &   0.07    & 1        &
\end{tabular}}
\end{table}

The convergence of the MCEM algorithm is slow, caused by irregularities of the likelihood function related to the identifiability problems described above.
If we fix the jump rate at its real value, we get the estimates $\hat{\rho}=1.019$, $\hat{\mu}=0.112$ and $\hat{\theta}=0.780$ (after just 7 iterations) and the estimated covariance matrix
{\scriptsize
\begin{center}
$\left(
  \begin{array}{cccc}
    0.0025367 &  0.00022448 &  -0.0011035 \\
   0.00022448 &   0.0015771 &  0.00059542 \\
   -0.0011035 &  0.00059542 &    0.038611 \\
  \end{array}
\right)$
\end{center}}

For the MCMC algorithm of Subsection \ref{idMCMCsec}, an EA step between observation, jump and refinement points is performed after each Barker's step. The chain is warmed-up for $5k$ iterations before the parameters start to be updated. The layer refinement idea described in Subsection \ref{impbkmcmc} is applied for $m=4$. The parameter vector is broken into four individual blocks. Parameters from the jump process $(\lambda,\theta)$ are sampled directly from their full conditionals and the other two parameters are sampled via Two-Coin Barker's. We adopt uniform prior distributions for $\mu$ and $\rho$ and the following informative priors: $\lambda\sim Exp(50)$ and $\theta\sim\mathcal{G}amma(7,6)$. As it is explained in Subsection \ref{mipi}, these informative priors play a crucial role in the identifiability of the model. They work in the direction of favouring a few larger jumps instead of many smaller jumps.

We set uniform random walk proposals for $\mu$ and $\rho$ having a higher acceptance rate then optimal but perform six alternate updates of each one on each iteration of the MCMC. Also, three consecutive updates of $X$ are performed on each iteration. Trace plots indicate good convergence. Posterior statistics are presented in Table \ref{tabsim6}. The chosen prior distributions seem to correct the irregularities of the likelihood and lead to reasonably good results.

\begin{table}[!h]
\caption{\label{tabsim6}Posterior statistics from the MCMC output.}
\centering
\fbox{\scriptsize
\begin{tabular}{c|c|c|c|c}
  & $\rho$ & $\mu$ & $\lambda$ & $\theta$ \\ \hline
  Mean          & 1.010 & 0.086 & 0.050 & 0.725 \\
  Median        & 1.010 & 0.086 & 0.046 & 0.709 \\
  Mode          & 1.013 & 0.086 & 0.039 & 0.667 \\
  St. Dev.      & 0.0719 & 0.051 & 0.025 & 0.189 \\
95\% Cred. Int. & (0.872,1.152) & (-0.014,0.188) & (0.013,0.111) & (0.396,1.131) \\ \hline
  real          & 1 & 0 & 0.07 & 1
\end{tabular}}
\end{table}


\section{Application}\label{appsect}

\subsection{Exchange rate USD$\times$GBP}

We consider the exchange rate between USD and GBP. The USD suffered a considerable depreciation during the 2008 world economic crisis. A few months later it had a moderate recovery and oscillated between 1.45 and 1.7 until beginning of 2016. We will use daily data from May 21, 2009, which was right after the recovery, up to March 27, 2013. This constitutes 1201 data points and is shown in Figure \ref{app1} in the log-scale. We shall adopt the MCMC procedure of Subsection \ref{MCMC_1}.

We choose to fit a scaled Brownian motion to model the continuous part of the process and proceed as follows to specify the jump part: 1. Plot time versus differences between consecutive log-observations. 2. Identify outlier differences as the values outside an interval that is defined by the (constant along time) limits of a dense cloud of points. 3. Consider the outlier values to be indicators of jumps in the process and use them to empirically estimate the jump rate and jump size distribution.
This analysis suggests a piecewise constant jump rate dividing the observed time interval in 5 parts and the absolute value of the jump sizes being well accommodated by a Gamma distribution, in particular, a $\mathcal{G}(8,1000)$. A mixture of a positive and a negative gamma distributions takes the probability mass of the jump size distribution away from zero and, consequently, helps to avoid identifiability problems. All of the empirical assumptions described in this paragraph are used only to elicit the model and are not actually imposed to the analysis. Naturally, some of the intervals could be likely to have more than one jump but, given the way the jump distribution is chosen and the fact that there are no restrictions in the number of jumps, this should happen only to a minority of the intervals estimated to have jumps and, therefore, not compromise the analysis.

The chosen model for the log-rate $V$ is given by:
\begin{eqnarray}\label{mdreal}
\ds dV_s &=& bdW_s+dJ_s,\;\;\;\;V_0=v,\;\;s\in[0,1201]  \\
  \lambda_1(s;V_{s-})&=&\lambda_i,\;\;\mbox{for }s\in A_i,\;i=1,\ldots,5\nonumber \\
  g_1(Z_j,V_{t_j-}) &=&  Z_j\sim p\mathcal{G}(8,1000)+(1-p)(-{\mathcal{G}}(8,1000)), \nonumber
\end{eqnarray}
for $A_1=[0,370]$, $A_2=(370,825]$, $A_3=(825,1000]$, $A_4=(1000,1120]$ and $A_5=(1120,1201]$.

We try to avoid identifiability problems by minimising the number of jumps and maximising $b$. This is done by setting informative priors to all the $\lambda_i$'s and by adopting the jump size distribution in (\ref{mdreal}). Parameters are assumed to be mutually independent with marginal priors: $b^2\sim\mathcal{U}(0,\infty),\;\;\lambda_i\sim\mathcal{G}(1,50)$, for all $i$, $p\sim\mathcal{U}(0,1)$.

It is enough to simulate only jump times and sizes in JBEA to derive the full conditional distributions of the parameters, which all have closed forms. Note however that, although this is a fairly simple model, exact inference is only feasible due to the JBEA algorithm.

We start the chain at values $b^2=0.002^2$, $\lambda_1=0.40$, $\lambda_2=0.28$, $\lambda_3=0.20$, $\lambda_4=0.07$, $\lambda_5=0.22$, $p=0.5$, and run 50$k$ iterations.
Standard diagnostics suggest that convergence is rapidly attained. Table \ref{pestreal} shows the posterior statistics of the parameters.

\begin{table}[!h]
\caption{\label{pestreal}Posterior statistics of the parameters for the last 40$k$ iterations.}
\centering
\fbox{\scriptsize
\begin{tabular}{c|ccccccc}

          & $b^2$ & $\lambda_1$ & $\lambda_2$ & $\lambda_3$ & $\lambda_4$ & $\lambda_5$ & $p$ \\ \hline
  Mean    & 0.000017  & 0.269 & 0.069 & 0.014 & 0.008 & 0.025 & 0.481 \\
  Median  & 0.000017  & 0.267 & 0.067 & 0.010 & 0.005 & 0.019 & 0.482 \\
  Std Dev & 0.0000014 & 0.046 & 0.029 & 0.013 & 0.008 & 0.022 & 0.059
\end{tabular}}
\end{table}

\subsection{S\&P500}

We now apply the methodology from Subsection \ref{MCEMudc} to fit the Pareto-Beta Jump-Diffusion (PBJD) to daily data from the S\&P500 index. The PBJD model was proposed in \citet{ramezani} to model stock price behavior. It allows for up and down jumps using a mixture jump size distribution to account for good and bad news. The model is the following:
\begin{eqnarray}\label{mdreal2}
\ds dV_s &=& \mu V_sds + \sigma V_sdW_s +dJ_s,\;\;\;\;V_0=v  \\
    \lambda_1(s;V_{s-})&=&\lambda,\;\;\;\;\; g_1(Z_j,V_{t_j-}) =  (Z_j-1)V_{t_j-} \nonumber \\
  Z_j&\sim&p\mbox{Pareto}(\eta_u)+(1-p)\mbox{Beta}(\eta_d,1), \nonumber
\end{eqnarray}
where the Pareto distribution takes values in $(1,\infty)$.

\begin{figure}[!h]
\centering \includegraphics[scale=0.25]{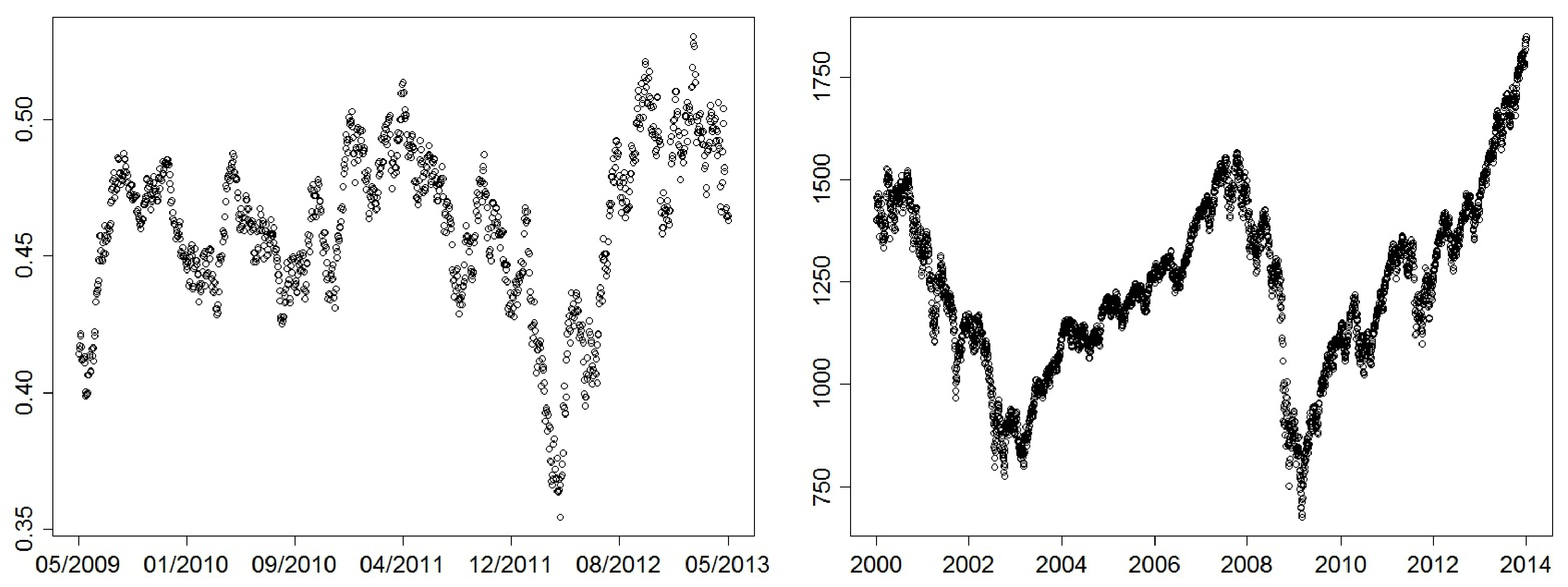}
\caption{Left: USD$\times$GBP log-rate, source: \emph{http://www.Investing.com}. Right: S\&P500 index, source: \emph{http://research.stlouisfed.org}}\label{app1}
\end{figure}

We consider daily data from 03/Jan/2000 to 31/Dec/2013 which consists of 3532 observations. This period incorporates the WTC 09/11 episode in 2001 and the 2008 economic crisis. The data is shown in Figure \ref{app1}. The two periods mentioned are clear in the graph and it seems reasonable to assume $\mu=0$, which considerably simplifies the algorithm. We apply the MCEM algorithm from Subsection \ref{ssecmcem}. Initial values are chosen through an empirical analysis of the data.
Results are presented in Table \ref{pestreal2}.

The estimated covariance matrix of the MLE returned a negative variance for parameter $p$. In order to resolve this we use a different parameterisation to compute this estimate. We make $\lambda_u=p\lambda$ and $\lambda_d=(1-p)\lambda$ - the rates of up and down jumps, respectively.
\begin{table}[!h]
\caption{\label{pestreal2}Maximum likelihood estimate and asymptotic 95\% confidence interval for the parameters.}
\centering
\fbox{\scriptsize
\begin{tabular}{c|ccccccc}
          & $\sigma$ & $\lambda$ & $p$ & $\lambda_u$ & $\lambda_d$ & $\eta_u$ & $\eta_d$ \\ \hline
  MLE        & 0.00577 & 0.841 & 0.532 & 0.447 & 0.394 & 144.5 & 125.2  \\
  C.I. 95\%  & (0.00497 , 0.00656) & - & - & (0.342 , 0.552) & (0.291 , 0.497) & (124.8 , 164.2) & (104.9 , 145.5)
\end{tabular}}
\end{table}

\section{Conclusions}

In this paper, we looked at the challenging problem of likelihood-based statistical inference for discretely observed jump-diffusion processes.
Our main contribution is to provide, for the first time, a comprehensive suite of methodologies for both (likelihood-based) frequentist and Bayesian {\em exact } inference for discretely observed jump-diffusion processes.

Our simplest methods are based on the JBEA algorithm for jump-diffusion bridge simulation and are presented in Section \ref{JBEAalgs}.  However more general methods are required for the case of unbounded drift and  jump-rate for the (transformed) process. To this end, we develop a more general approach  in Section \ref{inf2} which involve the construction of novel importance sampling and Barker MCMC  algorithms. Both these approaches are useful: the JBEA approach is far more computationally efficient, while the methods presented Section \ref{inf2} are more generally applicable.

We also address some important identifiability issues related to the inference problem and some practical implementation strategies. In particular, we propose (see Appendix E) an algorithm to simulate upper and lower bounds for a Brownian bridge and to simulate the bridge given these bounds and possibly other bridge points. We also concluded that prior distributions play a crucial role to solve identifiability problems. The Barker MCMC introduced in
Subsection \ref{idMCMCsec} seems also to be of independent interest.

The methods were empirically tested in some simulated examples and performed robustly and efficiently. The examples also addressed key identifiability issues. Finally, jump-diffusion models were used to fit some financial data.

Although beyond the scope of this paper, there is no intrinsic reason why this approach cannot be developed for many multi-dimensional contexts, at least for relatively small-dimensional models. However note that the exact simulation methodology that is used here, is fundamentally limited to reducible, gradient drift diffusions. Moreover the computational overheads associated with moderately high-dimensional problems are likely to be prohibitive.

Although we give the first general efficient methodology for exact likelihood-based inference for discretely observed jump-diffusions in this paper, we also acknowledge the restrictions and complexity involved in the methodology, which reflects the complexity of the inference problem. Most importantly though, we hope that our work will stimulate further work on our ambitious aim to solve the problem exactly.

\section*{Acknowledgements}

We would particularly like to thank the two anonymous referees who provided excellent and detailed comments on earlier versions of this paper.
Fl\'{a}vio Gon\c{c}alves
would like to thank FAPEMIG - grants PPM-00745-18 and APQ-01837-22, CNPq - grant 310433/2020-7 and the University of Warwick, for financial support. Krzysztof {\L}atuszy\'nski is supported by the Royal Society through the  Royal Society University Research Fellowship.  Gareth Roberts is supported by the EPSRC grants: {\em ilike}
(EP/K014463/1), CoSInES (EP/R034710/1) and Bayes for Health (EP/R018561/1).

\bibliographystyle{chicago}
\bibliography{biblio}

\section*{Appendix A - The JBEA algorithm}

Suppose that we want to simulate a jump-diffusion bridge from the process in expression (4) - Section 2, in the time interval $(0,t)$, conditioned to start at $x_{0}$ and end at $x_{t}$. The Lamperti transform implies that
\begin{equation}\label{alpha_g}
\ds \alpha(u)=\frac{b\{\eta^{-1}(u)\}}{\sigma\{\eta^{-1}(u)\}}-\frac{\sigma'\{\eta^{-1}(u)\}}{2}\;\;\;and\;\;\; g(z,u)=\eta(\eta^{-1}(u)+g_1(z,\eta^{-1}(u)))-u.
\end{equation}

Consider the probability measure $\mathbb{D}$ defined in Section 4.1 and restricted to the interval $(0,t)$.
Consider also the notation introduced in Section 2.1, so that $N_0$ and $(J_{0,1},\ldots,J_{0,N_0})$ are, respectively, the number of jumps and the jump sizes in $(0,t)$. Now define the probability measure $\mathbb{F}$ that, restricted to the interval $(0,t)$, satisfies
\begin{equation}\label{fF}
\ds \frac{d\mathbb{F}}{d\mathbb{D}}(X)= \frac{1}{c}\left(\exp\{\kappa_{0}\sum_{j=1}^{N_0}|J_{0,j}|\}\right),
\end{equation}
where $c$ and $\kappa_{0}$ are positive constants with the latter satisfying $|A(x)-A(y)|\leq \kappa_{0}|x-y|$, for all $x$ and $y$ in the state space of $X$, where $A(u):=\int_{0}^u\alpha(y)dy$. This means that $\mathbb{F}$ and $\mathbb{D}$ only differ in the distribution of $\ds(N_0,J_{0,1},\ldots,J_{0,N_0})$. In particular, the ratio $\frac{\pi_{\mathbb{F}}}{\pi_{\mathbb{D}}}(N_0,J_{0,1},\ldots,J_{0,N_0})$ of the densities of $(N_0,J_{0,1},\ldots,J_{0,N_0})$, under $\mathbb{F}$ and $\mathbb{D}$, is also given by the r.h.s. of (\ref{fF}). Details on how to simulate from $\pi_{\mathbb{F}}$ can be found in \citet{flavio2}.

Our aim is to retrospectively simulate bridges of the jump-diffusion with law $\mathbb{P}$ using rejection sampling with proposal law $\mathbb{F}$. The retrospective term effectively means that only a finite-dimensional (measurable) function of the diffusion path is simulated and this is enough to decide whether or not to accept the proposal. Additionally, once we have an acceptance, any other function of the diffusion path can be simulated directly from the proposal law, conditional on the already simulated function. The aforementioned finite-dimensional function includes: the jump times and sizes; a (simulable) function ${\textbf L_0}$ that defines upper and lower bounds for the process in $[0,t]$ (see Section 3.2 for a precise definition); the value of the process at the jump times and at another random collection of times used to simulate an auxiliary random variable (required to perform the accept/reject step - to be explained below).

Let $f_g(\cdot;s,X_{s-})$ be the jump size density of the target jump-diffusion at time $s$ and $\mathbb{P}|(x_0,x_t)$ be the probability measure of the jump-diffusion bridge in $(0,t)$ induced by $\mathbb{P}$. The following assumptions are required in order to simulate jump-diffusion bridges exactly using JBEA.
\renewcommand{\labelenumi}{(\alph{enumi})}
\emph{\begin{enumerate}\setlength\itemsep{0.1cm}
\item $\ds \alpha(\cdot)$ is continuously differentiable;
\item $\ds (\alpha^2+\alpha')(\cdot)$ is bounded below;
\item $\ds \mathbb{P}|(x,y)\ll\mathbb{D}$ and the Radon-Nikodym derivative of $\ds \mathbb{P}|(x,y)$ w.r.t. $\ds \mathbb{D}$ is given by (\ref{eqtgdj1}), for all $t\in\mathds{R}^+$ and for all $x$ and $y$ in the state space of $X$;
\item $\ds \lambda(s,u)$ is uniformly bounded for all $u$ in the state space of $X$ and all $s\in[0,T]$;
\item the probability measure induced by the density $f_g(\cdot;s,x)$, for all $s\in[0,T]$ and $x$ in the state space of $X$, is absolutely continuous w.r.t. the probability measure induced by the density $f_0$;
\item $\ds \frac{f_{g}(u;s,x)}{f_0(u)}$ is uniformly bounded for all $u\in E_g$ (support of $f_g$), $s\in[0,T]$ and $x$ in the state space of $X$;
\item $\lambda_0$ is chosen such that $\ds\frac{\lambda(s,x)f_{g}(u;s,x)}{\lambda_0 f_0(u)}\leq1$, for all  $u\in E_g$, $s\in[0,T]$ and $x$ in the state space of $X$;
\item $\ds A(u)$ is Lipschitz ($\alpha(\cdot)$ is uniformly bounded) and can be obtained analytically.
\item $\pi_{\mathbb{F}}$ is integrable and we can simulate from it.
\end{enumerate}}

One important property guaranteed by assumptions above is that the Radon-Nikodym (RN) derivative $\ds \frac{d\mathbb{P}}{d\mathbb{F}}(B|x_0,x_t)$ is bounded on the state space of $X$, implying that the rejection sampling algorithm that simulates from $\ds \mathbb{P}|(x_0,x_t)$ by proposing from $\ds \mathbb{F}$ is feasible. The acceptance probability $a(B)$ of this algorithm is proportion to that RN derivative and is given by
\begin{equation}\label{a.p.JBEA}
a(B) \eqdef \ds p_1(B)p_2(B)p_3(B),
\end{equation}
where
\begin{eqnarray}\label{phiD}
\ds &&p_1(B)=\exp\left\{-\frac{(x_t-J-x_0)^2}{2t}\right\}\;\;\; p_2(B)=\prod_{j=1}^{N}\left[\frac{\lambda(\tau_{j},B_{\tau_j-})f_{g}(J_{j};\tau_j,B_{\tau_j-})}
{\lambda_0 f_0(J_{j})}\exp\left\{-\Delta A(j)-\kappa_{0}|J_{j}|\right\}\right],\nonumber \\
&&p_3(B)=\exp\left\{-\ds\int_{0}^{t}\left(\phi(s,B_{s})-m\right)ds\right\};\nonumber\\
&&\phi(s,B_{s})=\left(\frac{\alpha^2+\alpha'}{2}\right)(B_{s})+\lambda(s,B_{s})\;\;,\;\;m=\inf_{u\in\mathds{R},0\leq s\leq t}\phi(s,u),\nonumber
\end{eqnarray}
with $\Delta A(j):=A(B_{\tau_{j}})-A(B_{\tau_{j}-})$.
\renewcommand{\labelenumi}{\arabic{enumi}.}

Let ${\textbf L_{0}}(B)$ be the proposed value of the variables ${\textbf L_0}$ and suppose that $r$ is a local (dependent on $B$) upper bound on $\phi (s,B_s)-m,\ 0\leq s\leq t$, which is obtained from the bounds on $B$ provided by ${\textbf L_{0}}(B)$. In order to decide whether to accept or not the proposal, we need to simulate three independent events of probabilities $p_1$, $p_2$ and $p_3$, respectively.

The first two are achieved by simulating independent Bernoulli r.v.'s (since $p_1$ and $p_2$ are known) through the simulation of two independent $\mathcal{U}(0,1)$ r.v.'s $U_{1}$ and $U_{2}$. For $i=1,2$, an event of probability $p_i$ occurs if $U_{i}\leq p_i$.

Now note that $p_3$ is equal to the probability that a Poisson process (PP) $\Phi$ with rate $r$ on $[0,t]\times[0,1]$
produces no points below the curve $(\phi (s,B_s)-m)/r$. This event can be evaluated by unveiling the proposal only at the time instances of the points from the PP, which makes it feasible to decide whether or not to accept the proposal. We call this the \emph{Poisson Coin} algorithm.

JBEA returns a finite-dimensional function (skeleton) of the jump-diffusion bridge with law $\ds \mathbb{P}|(x_0,x_t)$. This skeleton is loosely referred to as $\mathbf{S}$ in the main text of the paper. Any further points may be simulated from the proposal measure conditioned on the output skeleton.
Further details on the algorithm can be found in \citet{flavio2}, including extensions, simplifications, details on the simulation steps and practical strategies to optimise its computational cost.

\section*{Appendix B - Important results}

Let us recall some important notation and definitions previously introduced in the paper that shall be used to state the results in this appendix and present the proofs in Appendix C. For the jump-diffusion $X$ solving the SDE in (4) - Section 2, in the time interval $[t_{i-1},t_i]$, define ${\bf J}^{(i)}$ to be the jump times and sizes and $J_i$ to be the sum of the jump sizes, $X_{J}^{(i)}$ is the state of $X$ at the $N_i$ jump times and ${\textbf L_{i}}$ is a finite-dimensional function of $X$ that defines lower and upper bounds for $X$ in $[t_{i-1},t_i]$. Now, define also $\Phi^{(i)}$ to be the value of the Poisson process with rate $r_i(\theta)$ on $[t_{i-1},t_i]\times[0,1]$, which is used on the accept/reject step of JBEA and leads to an acceptance. Let $\kappa_i\sim Po\{r_i(\theta)\Delta t_i\}$ be the number of points from $\Phi^{(i)}$ with coordinates $\{\Psi^{(i)},\Upsilon^{(i)}\}=\{(\psi_{i,1},\upsilon_{i,1}),\ldots,(\psi_{i,\kappa_i},\upsilon_{i,\kappa_i})\}$ and define $\dot{X}^{(i)}$ to be the continuous bridges of $X$ between $t_{i-1}$, $t_i$ and jump times, linearly transformed to start and end at zero. Finally, let $U_{i,1}$ and $U_{i,2}$ be the two independent $U(0,1)$ random variables used to simulate the events of probability $p_1$ and $p_2$ appearing in the acceptance probability of JBEA in $[t_{i-1},t_i]$, as defined in Appendix A.

For $i=1,\ldots,n$ and $j=1,\ldots,N_i+1$, we define
\begin{equation}\label{ddotX}
\ds \ddot{X}_{s}^{(i,j)}=\ds \frac{1}{\sqrt{\tau_{i,j}-\tau_{i,j-1}}}\dot{X}_{\tau_{i,j-1}+s(\tau_{i,j}-\tau_{i,j-1})},\;\;\;s\in(0,1),
\end{equation}
and note that, under $\ds \mathbb{D}$, $\{\ddot{X}_{s}^{(i,j)};\tau_{i,j-1}<s<\tau_{i,j}\}$ is a standard Brownian bridge in $(0,1)$, whose measure we call $\mathbb{W}_{0,0}$. We then define $\ds \ddot{X}^{(i)}$ to be all the $\ds \ddot{X}_{s}^{(i,j)}$'s in $(t_{i-1},t_i)$ and $\ds \ddot{X}$ to be all the $\ds \ddot{X}^{(i)}$'s.

Finally, let $\mathbb{L}^{n}$ be the $n$-dimensional Lebesgue measure, $\mathbb{J}_{0,i}$ be the law of a jump process in $[t_{i-1},t_i]$ with unit jump rate and a parameter-free time and state-independent jump size density $f$ (same as in measure $\mathbb{Q}$) that defines a measure that dominates the jump size distributions in $\mathbb{P}$. Let also $\Phi_{i}^+$ be the measure of a unit rate Poisson process on $[t_{i-1},t_i]\times[0,1]$).

\begin{lemma}\label{POD2}
The conditional density of $\ds \left({\bf J}^{(i)},X_{J}^{(i)},\ddot{X}^{(i)},\Phi^{(i)},U_{i,1},U_{i,2}\right)$ given $\left(\theta,x_{i-1}(\theta),x_i(\theta)\right)$, under $\mathbb{P}$,
with respect to a dominating measure which is independent of $\theta$ is given by
\begin{eqnarray}\label{domephi}
\ds &&\pi\left({\bf J}^{(i)},X_{J}^{(i)},\ddot{X}^{(i)},\Phi^{(i)},U_{i,1},U_{i,2}|\theta,x_{i-1}(\theta),x_i(\theta)\right)\overset{\theta}{\propto}  \nonumber \\
&&\exp\left\{-(\Delta t_i)r_i(\theta)\right\}\left[r_i(\theta)\right]^{\kappa_i}
f_{\mathbb{F}}(N_i,J_{i,1},\ldots,J_{i,{N_i}};\theta)\pi_{\mathbb{Q}}(X_{J}^{(i)}|{\bf J}^{(i)},x_{i-1}(\theta),x_i(\theta);\theta) \nonumber\\
&&\mathds{1}\left[U_{i,1}\leq\exp\left\{-\frac{(\Delta x_i(\theta)-J_i)^2}{2\Delta t_i}\right\}\right] \nonumber \\
&&\mathds{1}\left[U_{i,2}\leq\prod_{j=1}^{N_i}\left[\frac{\ell(i,j;\theta)f_{g}(J_{i,j};i,j,\theta)}
{\lambda_i(\theta) f_i(J_{i,j};\theta)}\exp\left\{\left(\Delta A(i,j;\theta)\right)+\kappa_0(\theta)|J_{i,j}|\right\}\right]\right]\nonumber \\
&&\prod_{l=1}^{\kappa_i}\mathds{1}\Bigg[\frac{1}{r_i(\theta)}\dot{\phi}\Bigg(\psi_{i,l},\dot{X}_{\psi_{i,l}};\theta)
\Bigg)<\upsilon_{i,l}\Bigg] \frac{1}{a\left(x_{i-1}(\theta),x_i(\theta);\theta\right)},
\end{eqnarray}
where $\ds a\left(x_{i-1}(\theta),x_i(\theta);\theta\right)= E_{\mathbb{F}}\left[a(B;\theta)\right]$, for $a(B;\theta)$ as defined in (\ref{a.p.JBEA}). Also,\\
$\pi_{\mathbb{Q}}(X_{J}^{(i)}|{\bf J}^{(i)},x_{i-1}(\theta),x_i(\theta);\theta)$ is the density of $(X_{J}^{(i)}|{\bf J}^{(i)},x_{i-1}(\theta),x_i(\theta);\theta)$ under $\ds \mathbb{Q}$. Finally, $\kappa_0(\theta)$ is the constant $\kappa_0$ as defined in (\ref{fF}), for $(t,x_0,x_t)=(\Delta t_i,x_{i-1}(\theta),x_i(\theta))$.
\end{lemma}
\begin{thma}
The proof of this lemma is analogous to the proof of Lemma 3 from \citet{bpr06a} and, in our context, makes use of the results in Appendix G.
\end{thma}

\begin{lemma}\label{GRL2}
Consider the probability measure $\ds \mathbb{D}$ as defined in Subsection 4.1 and let $p_{\theta}(x_{i-1}(\theta),x_i(\theta);\Delta t_i)$ be the transition density of $X$ going from $x_{i-1}(\theta)$ at time $t_{i-1}$ to $x_i(\theta)$ at time $t_i$ under $\mathbb{P}$. Then
\begin{equation}\label{eqtgdj1}
\ds \frac{d\mathbb{P}}{d\mathbb{D}}({\bf J}^{(i)},X_{J}^{(i)},\dot{X}^{(i)}|x_{i-1}(\theta),x_i(\theta),\theta)=\mathcal{G}_i\frac{f_{N}(\Delta x_i(\theta)-J_i;0,\Delta t_i)}{p_{\theta}\left(x_{i-1}(\theta),x_i(\theta);\Delta t_i\right)},
\end{equation}
and
\begin{equation}\label{eqtgdj2}
\ds \frac{d\mathbb{P}}{d\mathbb{F}}({\bf J}^{(i)},X_{J}^{(i)},\dot{X}_i|x_{i-1}(\theta),x_i(\theta),\theta)=
\frac{d\mathbb{P}}{d\mathbb{D}}({\bf J}^{(i)},X_{J}^{(i)},\dot{X}_i|x_{i-1}(\theta),x_i(\theta),\theta)
c_i(\theta)\exp\left(-\kappa_0(\theta)\sum_{j=1}^{N_i}|J_{i,j}|\right),
\end{equation}
where $c_i(\theta)$ is the constant $c$ as defined in (\ref{fF}), for $(t,x_0,x_t)=(\Delta t_i,x_{i-1}(\theta),x_i(\theta))$, and
\begin{eqnarray}\label{GG}
\ds \mathcal{G}_i&=&\exp\left\{A(x_i(\theta);\theta)-A(x_{i-1}(\theta);\theta)
-\int_{t_{i-1}}^{t_i}\dot{\phi}(s,\dot{X}_s;\theta)ds+\Delta t_i\lambda_i(\theta) -\sum_{j=1}^{N_i}\Delta A(i,j;\theta)\right\} \nonumber \\
&\times&\prod_{j=1}^{N_i}\frac{\ell(i,j;\theta)f_g(J_{i,j};i,j,\theta)}{\lambda_i(\theta)f_i(J_{i,j};\theta)}.
\end{eqnarray}
\end{lemma}
\begin{thma}
See Lemma 2 and equation (22) from \citet{flavio2}.
\end{thma}

\begin{lemma}\label{GRL3}
\begin{equation}\label{eqtgdj3}
\ds \frac{d\mathbb{P}}{d\mathbb{Q}}(\dot{X}^{(i)}|{\bf J}^{(i)},X_{J}^{(i)},x_{i-1}(\theta),x_i(\theta),\theta)=\mathcal{G}_i\frac{f_{N}(\Delta x_i(\theta)-J_i;0,\Delta t_i)}{p_{\theta}\left(x_{i-1}(\theta),x_i(\theta);\Delta t_i\right)}
\bigg/\frac{d\mathbb{P}}{d\mathbb{Q}}({\bf J}^{(i)},X_{J}^{(i)}|x_{i-1}(\theta),x_i(\theta),\theta),
\end{equation}
for $\mathcal{G}_i$ as defined in Lemma \ref{GRL2}.
\end{lemma}
\begin{thma}
Use the same argument as in Lemma \ref{GRL2}.
\end{thma}

\begin{theorem}\label{MCMCJD}
The full conditional density of $\theta$ (w.r.t. the same dominating measure in $\pi(\theta)$) in the algorithm of Subsection 3.2 is given by:
\begin{eqnarray}\label{fullc2}
&& \pi\left(\theta|\mathbf{S},\textbf{v}\right) \propto \pi(\theta)\exp\left\{A(x_n(\theta);\theta)-A(x_0(\theta);\theta)\right\} \nonumber \\
&&
\prod_{i=1}^n\Bigg[\exp\left\{-\Delta t_i(r_i+m_i)(\theta)-\sum_{j=1}^{N_i}\left(\Delta A(i,j;\theta)\right)\right\} \left[r_i(\theta)\right]^{\kappa_i}
\left(\prod_{j=1}^{N_i}\ell(i,j;\theta)f_{g}(J_{i,j};i,j,\theta)\right) \left|\sigma(v_{i};\theta)^{-1}\right| \nonumber \\
&& f_{N}(X_{i,1-};x_{i-1}(\theta),\tau_{i,1}-t_{i-1}) f_{N}(x_i(\theta);X_{i,N_i},t_i-\tau_{i,{N_i}}) \prod_{l=1}^{\kappa_i}\left(1-\frac{\dot{\phi}\left(\psi_{i,l},\dot{X}_{\psi_{i,l}};\theta
\right)-m_i(\theta)}{r_i(\theta)}\right)\Bigg],
\end{eqnarray}
where
\begin{equation}\label{phifunc}
\ds m_i(\theta)=\inf_{u\in\mathcal{X},t_{i-1}\leq s\leq t_i}\phi(s,u;\theta)\;\;,\;\; r_i(\theta)=\sup_{u\in \bar{H}_i,t_{i-1}\leq s \leq t_i}\phi(s,u;\theta)-m_i(\theta),
\end{equation}
where $\mathcal{X}$ is the state space of $X$ and $\bar{H}_i$ is the interval defined by the local lower and upper bounds for $X_s$ in $[t_{i-1},t_i]$, which are obtained from ${\textbf L_{i}}$.
\end{theorem}
\begin{thma}
See Appendix C.
\end{thma}

\section*{Appendix C - proofs}

\subsection*{Proof of Lemma 1}

We shall find the density $\pi(\dot{V}_{com};\theta)$ of $\dot{V}_{com}=(\mathbf{v},\dot{V}_{mis})$, for $\dot{V}_{mis}=(\mathbf{J},X_J,\ddot{X})$, with respect to the product measure $\ds \mathbb{L}^{n}\otimes \mathbb{H}$, where $\mathbb{H}=\mathbb{H}_1\otimes\ldots\otimes\mathbb{H}_n$, $\ds \mathbb{H}_i=\sum_{k=0}^{\infty}\mathbb{H}_{k,i}$, for $i=1,\ldots,n$, and $\ds \mathbb{H}_{k,i}:=\mathbb{J}_{0,i}\otimes\mathbb{L}^k\otimes\mathbb{W}_{0,0}^{k+1}$, for $k=0,1,2,\ldots$.

We consider the following factorisation:

\begin{eqnarray}\label{dens_vcom1}
\ds \frac{d\mathbb{P}}{d(\mathbb{L}^{n}\otimes \mathbb{H})}(\dot{V}_{com}|\theta)&=&
\frac{d\mathbb{P}}{d\mathbb{L}^{n}}(\mathbf{v}|\theta)\frac{d\mathbb{P}}{d\mathbb{H}}(\mathbf{J},X_J,\ddot{X}|\mathbf{v},\theta)\nonumber\\
&=&\prod_{i=1}^n \left[ s_{\theta}(v_{i-1},v_i;\Delta t_i)  \frac{d\mathbb{P}}{d\mathbb{H}_i}(\mathbf{J}^{(i)},X_{J}^{(i)},\ddot{X}^{(i)}|\mathbf{v},\theta)\right], \nonumber
\end{eqnarray}
where $\ds s_{\theta}(a,b;\Delta t_i)$ is the transition density of $V$ going from $a$ at time $t_{i-1}$ to $b$ at time $t_i$ under the probability law defined as the solution for the SDE in (1) - Subsection 1.2.

We also have that
\begin{equation}\label{sp_q=equ}
\ds s_{\theta}(v_{i-1},v_i;\Delta t_i)= p_{\theta}(x_{i-1}(\theta),x_{i}(\theta);\Delta t_i)|\eta'(v_{t_i};\theta)|,
\end{equation}
for $p_{\theta}(x_{i-1}(\theta),x_{i}(\theta);\Delta t_i)$ as define in Lemma \ref{GRL2}.

The result in Proposition 5 in Appendix G (where a proof is provided) states that
\begin{equation}\label{ies5}
\ds \frac{d\mathbb{P}}{d\mathbb{H}_i}(\mathbf{J}^{(i)},X_{J}^{(i)},\ddot{X}^{(i)}|\mathbf{v},\theta)= \frac{d\mathbb{P}}{d\mathbb{Q}}(\mathbf{J}^{(i)},X_{J}^{(i)},\ddot{X}^{(i)}|\mathbf{v},\theta)\frac{d\mathbb{Q}}{d\mathbb{H}_i}(X_{J}^{(i)}|\mathbf{J}^{(i)},\mathbf{v},\theta).
\end{equation}
This and the fact that $\frac{d\mathbb{P}}{d\mathbb{Q}}(\ddot{X}^{(i)}|\mathbf{J}^{(i)},X_{J}^{(i)},\mathbf{v},\theta)=\frac{d\mathbb{P}}{d\mathbb{Q}}(\dot{X}^{(i)}|\mathbf{J}^{(i)},X_{J}^{(i)},\mathbf{v},\theta)$ imply that
\begin{equation}\label{dens_vcom2}
\ds \frac{d\mathbb{P}}{d\mathbb{H}_i}(\mathbf{J}^{(i)},X_{J}^{(i)},\ddot{X}^{(i)}|\mathbf{v},\theta)
=\frac{d\mathbb{P}}{d\mathbb{Q}}(\mathbf{J}^{(i)},X_{J}^{(i)}|\mathbf{v},\theta)\frac{d\mathbb{P}}{d\mathbb{Q}}(\dot{X}^{(i)}|\mathbf{J}^{(i)},X_{J}^{(i)},\mathbf{v},\theta)
\frac{d\mathbb{Q}}{d\mathbb{H}_i}(X_{J}^{(i)}|\mathbf{J}^{(i)},\mathbf{v},\theta).
\end{equation}

Finally, note that
\begin{eqnarray}\label{dens_vcom3}
\ds \frac{d\mathbb{Q}}{d\mathbb{H}_i}(X_{J}^{(i)}|\mathbf{J}^{(i)},\mathbf{v},\theta)&=&\frac{d\mathbb{Q}}{d\mathbb{H}_i}(X_{J}^{(i)}|\mathbf{J}^{(i)},x_{i-1}(\theta),x_i(\theta),\theta)=
\frac{\ds \frac{d\mathbb{Q}}{d(\mathbb{H}_i\otimes\mathbb{L})}\left(X_{J}^{(i)},x_i(\theta)|{\bf J}^{(i)},x_{i-1}(\theta),\theta\right)}
{\frac{d\mathbb{Q}}{d\mathbb{L}}\left(x_i(\theta)|{\bf J}^{(i)},x_{i-1}(\theta),\theta\right)} \nonumber \\
&\stackrel{\theta}{\propto}& \frac{f_{N}(X_{i,1-};x_{i-1}(\theta),\tau_{i,1}-t_{i-1}) f_{N}(x_i(\theta);X_{i,N_i},t_i-\tau_{i,{N_i}})}{f_{N}(\Delta x_i(\theta)-J_i;0,\Delta t_i)}.
\end{eqnarray}

Now substitute (\ref{dens_vcom3}) and (\ref{eqtgdj3}) into (\ref{dens_vcom2}), then substitute (\ref{dens_vcom2}) and (\ref{sp_q=equ}) into (\ref{dens_vcom1}) and note that $\eta'(u;\theta)=\sigma(u;\theta)^{-1}$.

\begin{flushright}
$\square$
\end{flushright}

\subsection*{Proof of Theorem \ref{MCMCJD}}

We use the fact that $\pi\left(\theta,{\bf J},X_{J},\dot{X},\Phi|\textbf{v}\right)\overset{\theta}{\propto}\pi\left(\theta,{\bf J},X_{J},\ddot{X},\Phi|\textbf{v}\right)$ and consider the following decomposition:
\begin{equation}\label{eqtful}
\ds \pi\left(\theta,{\bf J},X_{J},\ddot{X},\Phi|\textbf{v}\right)=\pi(\theta|\textbf{v})\pi\left({\bf J},X_{J},\ddot{X},\Phi|\theta,\textbf{v}\right)=
\pi(\theta|\textbf{v})\prod_{i=1}^n\pi\left({\bf J}^{(i)},X_{J}^{(i)},\ddot{X}^{(i)},\Phi^{(i)}|\theta,\textbf{v}\right),
\end{equation}
where the second equality is obtained by the Markov property. Now note that
\begin{equation}\label{eqtth}
\ds \pi(\theta|\textbf{v})\propto\pi(\theta)\prod_{i=1}^np_{\theta}\left(x_{i-1}(\theta),x_i(\theta);\Delta t_i\right)\left|\eta'(v_i;\theta)\right|.
\end{equation}
We now obtain an equality for the transition density $\ds p_{\theta}\left(x_{i-1}(\theta),x_i(\theta);\Delta t_i\right)$ to be substituted into (\ref{eqtth}).
First, we substitute (\ref{GG}) into (\ref{eqtgdj1}) and then (\ref{eqtgdj1})  in (\ref{eqtgdj2}) and take the expectation with respect to $\ds \mathbb{F}$ on both sides of (\ref{eqtgdj2}) to get
\begin{equation}\label{eqtd}
\ds p_{\theta}\left(x_{i-1}(\theta),x_i(\theta);\Delta t_i\right)=\mathbb{E}_{\mathbb{F}}\left[
\mathcal{G}_if_{N}(\Delta x_i(\theta)-J_i;0,\Delta t_i)
\frac{f_{\mathbb{D}}}{f_{\mathbb{F}}}(N_i,J_{i,1},\ldots,J_{i,{N_i}})\right].
\end{equation}
The last term inside the expectation is given in (\ref{fF}). The result in (\ref{eqtd}) comes from the fact that the expectation of the l.h.s. of (\ref{eqtgdj2}) is 1 and the term $\ds p_{\theta}\left(x_{i-1}(\theta),x_i(\theta);\Delta t_i\right)$ is a constant w.r.t. the expectation.

Second, we take the expectation of the acceptance probability of JBEA in (\ref{a.p.JBEA}) w.r.t. $\ds \mathbb{F}$ to obtain the global acceptance probability $\ds a\left(x_{i-1}(\theta),x_i(\theta);\theta\right)$ of JBEA.
\begin{eqnarray}\label{reca}
&&a(x_{i-1}(\theta),x_i(\theta);\theta)=\mathbb{E}_{\mathbb{F}}\Bigg[\exp\Bigg\{-\int_{t_{i-1}}^{t_i}
\dot{\phi}(s,\dot{X}_s;\theta)-m_i(\theta)ds \nonumber \\
&&-\sum_{j=1}^{N_i}\left(\Delta A(i,j;\theta)+\kappa_0(\theta)|J_{i,j}|\right)\Bigg\}
\prod_{j=1}^{N}\frac{\lambda(\cdot j;\theta)f_g(J_{i,j};j,\theta)}{\lambda_i(\theta)f_i(J_{i,j};\theta)}\exp\left\{-\frac{1}{2\Delta t_i}(\Delta x_i(\theta)-J_i)^2\right\}\Bigg].
\end{eqnarray}
Now, comparing (\ref{reca}) with (\ref{eqtd}) and using the result in (\ref{fF}), we get
\begin{eqnarray}\label{eqden}
\ds p_{\theta}\left(x_{i-1}(\theta),x_i(\theta);\Delta t_i\right)&\propto&\exp\left\{A(x_i(\theta);\theta)-A(x_{i-1}(\theta);\theta)-
\Delta t_i(m_i(\theta)-\lambda_i(\theta))\right\} \nonumber \\
&&c_i(\theta)a(x_{i-1}(\theta),x_i(\theta);\theta).
\end{eqnarray}

Finally, the desired result is obtained by: i) substituting (\ref{eqden}) into (\ref{eqtth}) and noting that $\eta'(u;\theta)=\sigma(u;\theta)^{-1}$; ii) integrating out the $U_{i,1}$'s, $U_{i,2}$'s and $\upsilon_{i,l}$'s in (\ref{domephi}); iii) substituting (\ref{domephi}) and (\ref{eqtth}) into (\ref{eqtful}).

\begin{flushright}
$\square$
\end{flushright}

\subsection*{Proof of Proposition 1}

We have that
\begin{eqnarray*}
\ds\mathbb{E}_{\mathbb{P}|\mathbf{x}}\left[l(\theta)\right]   &=&
\mathbb{E}_{\mathbb{P}|\mathbf{x}}\left[\sum_{i=1}^nl_i(\theta)\right] = \sum_{i=1}^n\mathbb{E}_{\mathbb{P}|\mathbf{x}}\left[l_i(\theta)\right]
=\sum_{i=1}^n\mathbb{E}_{\mathbb{D}}\left[wl_i(\theta)\right]=
\sum_{i=1}^n\mathbb{E}_{\mathbb{D}}\left[\prod_{k=1}^nw_{k}l_i(\theta)\right] \\
   &=&  \sum_{i=1}^n\prod_{k\neq i}\mathbb{E}_{\mathbb{D}}\left[w_{k}\right]\mathbb{E}_{\mathbb{D}}\left[w_{i}l_i(\theta)\right]
=\sum_{i=1}^n\mathbb{E}_{\mathbb{D}}\left[w_il_i(\theta)\right],
\end{eqnarray*}
where the fifth equality is justified by the Markov property and the last equality by the fact that $\ds \mathbb{E}_{\mathbb{F}}\left[w_i\right]=1,\;\forall\;i$.
\begin{flushright}
$\square$
\end{flushright}

\subsection*{Proof of Proposition 2}

Firstly, note that
$\ds Var[E_1]=\frac{1}{M}Var\left[\sum_{i=1}^n{\left(\prod_{k=1}^nw_k\right)l_i(\theta)}\right]$ and
$\ds Var[E_2]=\frac{1}{M}Var\left[\sum_{i=1}^nw_{i}l_i(\theta)\right]$.
Moreover, $w_i \independent w_j,\;\forall i\neq j$ and $\mathbb{E}_{\mathbb{F}}[w_i]=1$.

Define $A_i=\prod_{k\neq i}w_k$ and $A_{ij}=\prod_{k\neq i,j}w_k$. Defining $l_i=l_i(\theta)$, we have that
\begin{eqnarray}
\ds M.Var[E_1]&=&Var\left[\sum_{i=1}^nA_iw_il_i\right]=\mathbb{E}\left[\left(\sum_{i=1}^nA_iw_il_i\right)^2\right]-\left(\mathbb{E}\left[\sum_{i=1}^nA_iw_il_i\right]\right)^2 \nonumber\\
&=&\sum_{i=1}^n\mathbb{E}[A_{i}^2]\mathbb{E}[(w_il_i)^2]+2\sum_{i<j}\mathbb{E}[A_{ij}]\mathbb{E}[w_il_i]\mathbb{E}[w_jl_j]-\left(\sum_{i=1}^n\mathbb{E}[A_i]\mathbb{E}[w_il_i]\right)^2 \nonumber \\
&\geq& \sum_{i=1}^n\mathbb{E}[(w_il_i)^2]+2\sum_{i<j}\mathbb{E}[w_il_i]\mathbb{E}[w_jl_j]-\left(\sum_{i=1}^n\mathbb{E}[w_il_i]\right)^2 \nonumber \\
&=& \mathbb{E}\left[\left(\sum_{i=1}^nw_il_i\right)^2\right]-\left(\mathbb{E}\left[\sum_{i=1}^nw_il_i\right]\right)^2=M.Var[E_2],
\end{eqnarray}
where all the expectations are w.r.t. $\mathbb{D}$.

\subsection*{Proof of Proposition 3}

Firstly, note that
\begin{equation}\label{proofIC1}
\ds \mathbb{E}_{\dot{U}}\left[ \Delta t_i \phi(\dot{U}_i,X_{\dot{U}_i};\theta) \right] =
\left[- \int_{t_{i-1}}^{t_i}\phi(s,X_{s-};\theta)ds \right]
\end{equation}

Now, we suppress the notation $\theta'$ from $\mu_i$, $b_i$ and $\phi$ and show that
\begin{equation}\label{proofIC2}
\ds \mathbb{E}_{K,U}\left[ e^{(\mu_i(X)-b_i(X))\Delta t_i}\mu_{i}(X)^{-K_i} \prod_{k=1}^{K_i}\left(b_i(X)-\phi\left(U_{i,k},X_{U_{i,k}}\right)\right) \right] =  \left[ \exp\left\{-\int_{t_{i-1}}^{t_i}\phi(s,X_{s})ds\right\} \right]
\end{equation}

We set $b_i=b_i(X)$ and $\mu_i=\mu_i(X)$ for cleanness of notation.
\begin{eqnarray*}
\ds
 && \mathbb{E}_{K,U}\left[ e^{(\mu_i-b_i)\Delta t_i}\mu_{i}^{-K_i}\prod_{k=1}^{K_i}\left(b_i-\phi\left(U_{i,k},X_{U_{i,k}}\right)\right) \right] \\
 &=& e^{(\mu_i-b_i)\Delta t_i} \mathbb{E}_{K}\left[ \mu_{i}^{-K_i} \prod_{k=1}^{K_i}\left(b_i- \frac{1}{\Delta t_i}\int_{t_{i-1}}^{t_i}\phi(s,X_{s})ds \right) \right]  \\
 &=& e^{(\mu_i-b_i)\Delta t_i} \mathbb{E}_{K}\left[ (\Delta t_i\mu_{i})^{-K_i} \left(\Delta t_ib_i- \int_{t_{i-1}}^{t_i}\phi(s,X_{s})ds \right)^{K_i} \right] \\
 &=&  e^{(\mu_i-b_i)\Delta t_i} e^{-\mu_i\Delta t_i}
 \sum_{K_i=0}^{\infty}\frac{\left(\Delta t_ib_i- \int_{t_{i-1}}^{t_i}\phi(s,X_{s})ds \right)^{K_i}}{K_i!} =
 \exp\left\{ -\int_{t_{i-1}}^{t_i}\phi(s,X_{s})ds \right\}
\end{eqnarray*}
The desired result is obtained by replacing the r.h.s. of (\ref{proofIC1}) and (\ref{proofIC2}) by the respective l.h.s. in the expression of $w_il_i(\theta)$.

\subsection*{Proof of Lemma 2}

To prove that the estimator is strongly consistent, note that
\begin{equation*}
\ds Y_{M}^{(i1)}=\frac{1}{M}\sum_{k=1}^M\left(\dot{w}_{i}^{(k)}\dot{l}_{i}^{(k)}(\theta)\right)\rightarrow
\mathbb{E}_{\mathbb{P}|\mathbf{x}}[l_i(\theta)],\;\;\;
\mbox{a.s. as }M\rightarrow\infty,\mbox{ by SLLN},
\end{equation*}
\begin{equation*}
\ds Y_{M}^{(i2)}=\frac{1}{M}\sum_{k=1}^M\left(\dot{w}_{i}^{(k)}\right)\rightarrow1,\;\;\;
\mbox{a.s. as }M\rightarrow\infty,\mbox{ by SLLN}.
\end{equation*}
Since $Y_{M}^{(i1)}/Y_{M}^{(i2)}$ is a a.s. continuous function of $(Y_{M}^{(i1)},Y_{M}^{(i2)})$, the result is established by the Convergence of Transformations Theorem.

To prove the finite variance part, we write $\dot{l}_i$ instead of $\dot{l}_i(\theta)$ and note that $\ds {\tilde{w}_{i}}^{(k)}<1\;\;a.s.$ and
\begin{eqnarray}
\ds Var[\dot{E}_3]&=&
Var\left[\sum_{i=1}^n\sum_{k=1}^M\left({\tilde{w}_{i}}^{(k)}\dot{l}_{i}^{(k)}\right)\right]=
\sum_{i=1}^nVar\left[\sum_{k=1}^M\left({\tilde{w}_{i}}^{(k)}\dot{l}_{i}^{(k)}\right)\right] \nonumber \\
&=&\sum_{i=1}^n\left(\sum_{k=1}^MVar\left[{\tilde{w}_{i}}^{(k)}\dot{l}_{i}^{(k)}\right]
+2\sum_{1\leq k_1<k_2\leq M}Cov\left({\tilde{w}_{i}}^{(k_1)}\dot{l}_{i}^{(k_1)},{\tilde{w}_{i}}^{(k_2)}\dot{l}_{i}^{(k_2)}\right)\right). \nonumber
\end{eqnarray}
Therefore,
$$\ds Var[\dot{E}_3]<\infty\Leftrightarrow Var\left[{\tilde{w}_{i}}^{(k)}\dot{l}_{i}^{(k)}\right]<\infty
\mbox{ and }|Cov\left({\tilde{w}_{i}}^{(k_1)}\dot{l}_{i}^{(k_1)},{\tilde{w}_{i}}^{(k_2)}\dot{l}_{i}^{(k_2)}\right)|<\infty,\;\forall\;i,k,k_1,k_2.$$
Finally,
$$\mathbb{E}\left[\left({\tilde{w}_{i}}^{(k)}\dot{l}_{i}^{(k)}\right)^2\right]\leq
\mathbb{E}\left[\left(\dot{l}_{i}^{(k)}\right)^2\right],\;\forall\;i,k,$$
\begin{eqnarray}
\ds &&|Cov\left({\tilde{w}_{i}}^{(k_1)}\dot{l}_{i}^{(k_1)},{\tilde{w}_{i}}^{(k_2)}\dot{l}_{i}^{(k_2)}\right)|=
|\mathbb{E}\left[{\tilde{w}_{i}}^{(k_1)}{\tilde{w}_{i}}^{(k_2)}\dot{l}_{i}^{(k_1)}\dot{l}_{i}^{(k_2)}\right]-
\mathbb{E}\left[{\tilde{w}_{i}}^{(k_1)}\dot{l}_{i}^{(k_1)}\right]\mathbb{E}\left[{\tilde{w}_{i}}^{(k_2)}\dot{l}_{i}^{(k_2)}\right]| \nonumber \\
&\leq&\mathbb{E}\left[|\dot{l}_{i}^{(k_1)}\dot{l}_{i}^{(k_2)}|\right]+
\mathbb{E}\left[|\dot{l}_{i}^{(k_1)}|\right]\mathbb{E}\left[|\dot{l}_{i}^{(k_2)}|\right]=2\left(\mathbb{E}\left[|\dot{l}_{i}^{(k)}|\right]\right)^2\leq
2\mathbb{E}\left[\left(\dot{l}_{i}^{(k)}\right)^2\right] \;\forall\;i,k,k_1,k_2.\nonumber
\end{eqnarray}
\begin{flushright}
$\square$
\end{flushright}

\section*{Appendix D - Estimator for the covariance matrix of the MLE}

The covariance matrix of the MLE is approximated by the inverse of the observed information matrix $I$, that is
\begin{equation}
\ds I^{-1}=\left[-\frac{\partial^2}{\partial\theta^2}l(\mathbf{x};\theta)\right]^{-1},
\end{equation}
given that the required regularity conditions are satisfied. Now Oakes' identity \citep[see][]{oakes} can be rewritten to give
\begin{eqnarray}
\ds \frac{\partial^2}{\partial\theta^2}l(\mathbf{x};\theta)&=&
\mathbb{E}_{X_{mis}^{\tilde{\mathbb{P}}}|x_{obs},\theta'}\left[\frac{\partial^2}{\partial\theta^2}l(X_{com};\theta)\right]+
\mathbb{E}_{X_{mis}^{\tilde{\mathbb{P}}}|x_{obs},\theta'}\left[\left(\frac{\partial}{\partial\theta}l(X_{com};\theta)\right)^2\right] \nonumber \\
&&-
\left[\mathbb{E}_{X_{mis}^{\tilde{\mathbb{P}}}|x_{obs},\theta'}\left(\frac{\partial}{\partial\theta}l(X_{com};\theta)\right)\right]^2. \nonumber
\end{eqnarray}
We can use the general result from Proposition 1 to obtain
\begin{equation}\label{emvvar}
\ds\left.\frac{\partial^2}{\partial\theta^2}l(\mathbf{x};\theta)\approx(M_1+M_2+M_3-M_4)\right|_{\theta=\hat{\theta}},
\end{equation}
where $\ds \hat{\theta}$ is the MLE of $\theta$ and each $M$ is a matrix with the $[r,c]$-th entry given by:
\begin{eqnarray}
\ds M_1[r,c]&=& \sum_{i=1}^n\sum_{k=1}^M {\tilde{w}_{i}}^{(k)}\frac{\partial^2}{\partial\theta_r\partial\theta_c}\dot{l}_{i}^{(k)};\;\;\; M_2[r,c]= \sum_{i=1}^n \mathbb{E}_{\mathbb{P}}\left[\frac{\partial}{\partial\theta_r}l_{i}\frac{\partial}{\partial\theta_c}l_{i}\right]; \nonumber \\
\ds M_3[r,c]&=& \sum_{i_1=1}^n\sum_{i_2\neq i_1}\left(\sum_{k=1}^M {\tilde{w}_{i}}^{(k)}\frac{\partial}{\partial\theta_r}\dot{l}_{i_1}^{(k)}\right)
\left(\sum_{k=1}^M {\tilde{w}_{i}}^{(k)}\frac{\partial}{\partial\theta_c}\dot{l}_{i_2}^{(k)}\right); \nonumber \\
\ds M_4[r,c]&=& \left(\sum_{i=1}^n\sum_{k=1}^M {\tilde{w}_{i}}^{(k)}\frac{\partial}{\partial\theta_r}\dot{l}_{i}^{(k)}\right)
\left(\sum_{i=1}^n\sum_{k=1}^M {\tilde{w}_{i}}^{(k)}\frac{\partial}{\partial\theta_c}\dot{l}_{i}^{(k)}\right).\nonumber \\
\end{eqnarray}
To obtain the equations above we use the Leibniz Integral Rule which allows us to exchange the partial derivative w.r.t. $\theta$ and the integral in $ds$ - that is because $X(s)$ is \emph{c\`{a}dl\`{a}g}.
To deal with $M_2$, note that double integrals will emerge from the product and we use the following identity to obtain the MC estimator:
\begin{equation}\label{mcedi}
\ds \int_{t_{i-1}}^{t_i}\int_{t_{i-1}}^{t_i}f_{1}^*(X(s_1),\theta)f_{2}^*(X(s_2),\theta)ds_1ds_2=
(\Delta t_i)^2\mathbb{E}_{\dot{U}_{i,1},\dot{U}_{i,2}}\left[f_{1}^*(X(\dot{U}_{i,1}),\theta)f_{2}^*(X(\dot{U}_{i,2}),\theta)\right],
\end{equation}
where $\dot{U}_{i,1}$ and $\dot{U}_{i,2}$ are independent and uniformly distributed in $[t_{i-1},t_i]$.

Finally, the same algorithm may be used to estimate the covariance matrix of the MLE when using the algorithms from Subsection 2.1.
In that case, we have $\tilde{w}_{i}^{(k)}=1/M$, which leads to further simplifications in the formule.

\section*{Appendix E - The layered Brownian bridge}

The MCEM and MCMC algorithms presented in Sections 3 and 4 require the simulation of lower and upper bounds for a collection of Brownian bridges.
This can be done by simulating a finite-dimensional random variable which is referred to as ${\textbf L_{i}}$ throughout the paper. The algorithm to do so is called the layered Brownian bridge and basically samples layers that contains the supremum and infimum of the bridges.
The algorithm to construct and simulate the layered Brownian bridge (layers and bridge points given the layers) was introduced in \citet{bpr07}.

Equally important, the algorithms from Subsection 4.2 also require the simulation of extra points of the bridge given the layers and other points previously simulated. This is not a trivial task and no explicit algorithm to do so is presented in the original paper. For that reason we present here an overview of the layered Brownian bridge algorithm from \citet{bpr07} and devise an algorithm to perform the simulation of the extra points.

In fact, we restrict ourselves to the simulation of standard bridges (starting and ending in 0) as this is necessary when there are unknown parameters in the diffusion coefficient and also offers a more efficient (tighter bounds) solution in any case. We also restrict ourselves to the symmetric layer case as this is much simpler and there is no real advantage in using asymmetric layers.

Suppose, without loss of generality, a standard Brownian bridge $W$ in $[0,t]$. Let $\{b_i\}_{i\geq1}$ be an increasing sequence of positive real numbers with $b_0=0$ and define the following events:
$$\overline{D}_k=\left\{\ds \sup_{0 \leq s \leq t}W_s\in [b_{k-1},b_k]\right\}\bigcap\left\{\ds \inf_{0 \leq s \leq t}W_s>-b_k\right\},$$
$$\underline{D}_k=\left\{\ds \inf_{0 \leq s \leq t}W_s\in [-b_k,-b_{k-1}]\right\}\bigcap\left\{\ds \sup_{0 \leq s \leq t}W_s<b_k\right\},$$
$$D_k=\overline{D}_k\bigcup \underline{D}_k,\;k\geq1.$$

Define the random variable $I=I(W)$ such that $\{I=k\}=D_k$ and note that $\{I=k\}$ implies that $\{-b_k<W_s<+b_k,\;\forall\;s\;\in [0,t]\}$.
\citet{bpr07} show that
\begin{equation}\label{ea3-1}
\ds    F(k):=P(I\leq k)=\gamma\left(t;0,0,b_k\right),
\end{equation}
for $i\geq 1$, where
$$\ds    \gamma\left(s,u_1,u_2,K\right)=1-\sum_{j=1}^{\infty}\left\{\sigma_j(s,u_1,u_2,K)-\tau_j(s,u_1,u_2,K)\right\},$$
$$\ds    \sigma_j\left(s,u_1,u_2,K\right)=\bar{\sigma}_j(s,u_1,u_2,K)+\bar{\sigma}_j(s,-u_1,-u_2,K),$$
$$\ds    \tau_j\left(s,u_1,u_2,K\right)=\bar{\tau}_j(s,u_1,u_2,K)+\bar{\tau}_j(s,-u_1,-u_2,K),$$
$$\ds    \bar{\sigma}_j\left(s,u_1,u_2,K\right)=\exp\left\{-\frac{2}{s}[2Kj-(K+u_1)][2Kj-(K+u_2)]\right\},$$
$$\ds    \bar{\tau}_j\left(s,u_1,u_2,K\right)=\exp\left\{-\frac{2j}{s}[4K^2j+2K(u_1-u_2)]\right\}.$$

We sample $I$ by using the inverse c.d.f. method and, although we cannot evaluate $F(k)$ exactly, we can find bounds for it and apply the alternating series method as follows.

Define $\ds S_{2j}^{(k)}=1-\sum_{l=1}^{j-1}(\sigma_l-\tau_l)-\sigma_j$ and $\ds S_{2j-1}^{(k)}=1-\sum_{l=1}^{j}(\sigma_l-\tau_l)$, for $j\geq1$, with $\sigma$ and $\tau$ defined for $I=k$.
It can be shown that
\begin{equation}\label{ea3-6}
\ds   0<S_{2}^{(k)}<S_{4}^{(k)}<S_{6}^{(k)}<\ldots<F(i)<\ldots<S_{5}^{(k)}<S_{3}^{(k)}<S_{1}^{(k)},
\end{equation}
with $S_{2j}^{(k)}\uparrow F(k)$ and $S_{2j-1}^{(k)}\downarrow F(k)$. Defining $S_{j}^{(0)}=0$, $\forall\;j\geq1$, if $u\sim U(0,1)$,  we have the following result:
\begin{equation}\label{ea3-7}
\ds   I=k,\;\;\mbox{if } S_{2j-1}^{(k-1)}<u<S_{2j}^{(k)},\;\;\mbox{for some } j\geq1.
\end{equation}

This way, we can use the following algorithm to simulate $I$:\\
\\
\begin{tabular}[!]{|l|}
\hline\\
\parbox[!]{13cm}{
\texttt{
{\bf Algorithm to simulate $I$}
\begin{enumerate}
  \item Simulate $u\sim U(0,1)$;
  \item make $k=1$;
  \item make $j=1$;
  \item compute $S_{2j-1}^{(k)}$ and $S_{2j}^{(k)}$;
\begin{enumerate}
  \item if $u<S_{2j}^{(k)}$, output $I=k$;
  \item if $u>S_{2j-1}^{(k)}$, make $i=k+1$, $j=1$ and GOTO 4;
  \item else, make $j=j+1$ and GOTO 4.
\end{enumerate}
\end{enumerate}
}}\\  \hline
\end{tabular}\\

Now that we have the algorithm to sample the layers, we shall move on to the algorithm to simulate from $W_s|I=k$. In the MCMC algorithm from Subsection 4.2, this simulation will be required at different stages of the algorithm and may fall into one of the two cases: $i)$ no points from the bridge have yet been simulated; $ii)$ some points of the bridge have already been simulated; which in turn fall into two more cases: $a)$ $I=1$; $b)$ $I\geq2$.
A nice solution for the case $ia$ (and implicitly for case $iia$) is presented in \citet{bpr07}. The authors also present an algorithm for $ib$, but not for $iib$. The latter is more complex and requires some work to devise an efficient solution. We propose the following general strategy, which shall solve the problem in any of the four cases described above.

The first step after $I$ is simulated is to simulate the extreme (minimum or maximum) that reaches the most external layer - note that $I=k$ implies that either the minimum is in $(-b_k,-b_{k-1})$ or the maximum is in $(b_{k-1},b_k)$. This is done via rejection sampling with proposal
\begin{equation}\label{ea3-9}
\ds   \mathbb{P}_{D_I}:=\frac{1}{2}BB_{\overline{M}_I}+\frac{1}{2}BB_{\underline{M}_I},
\end{equation}
where $\ds BB_{\overline{M}_I}$ is the Brownian bridge $W$ conditional on $\ds \overline{M}_I:=\max_{0\leq s\leq t}\{W_s\}\in[b_{I-1},b_I]$ and
$\ds BB_{\underline{M}_I}$ is $W$ conditional on $\ds \underline{M}_I:=\min_{0\leq s\leq t}\{W_s\}\in[-b_I,-b_{I-1}]$.

To simulate from this proposal, we first choose between $\ds BB_{\overline{M}_I}$ and $\ds BB_{\underline{M}_I}$ with probability 1/2 each.
If we choose the first (second) one, we have to simulate the maximum (minimum) of $W$ given that it is in the given interval.

Let $m$ be the minimum and $M$ be the maximum of $W$ in $[0,t]$. In order to simulate the minimum conditional on being in an interval $[m_1,m_2]$, we need to compute
the distribution function $F_m$ of this minimum, which is:
\begin{equation}\label{ea3-10}
\ds   F_m(w)=\exp\left\{\frac{-2w^2}{t}\right\}.
\end{equation}

Let $\ds U_1\sim U[F(m_1),F(m_2)]$ and $\ds E=-\log(U_1)$, then
\begin{equation}\label{ea3-11}
\ds   m=\frac{\left(-\sqrt{2tE}\right)}{2}.
\end{equation}

In order to simulate the last time instant $t_m$ where this minimum is attached, set $\ds c_1=\frac{m^2}{2t}$ and let $\ds U_2\sim U(0,1)$, $\ds I_1\sim IGau\left(1,2c_1\right)$ and $\ds I_2\sim 1/IGau\left(1,2c_1\right)$ and define
$\ds V=\mathbb{I}\left[U_2<1/2\right]\cdot I_1 + \mathbb{I}\left[U_2\geq1/2\right]\cdot I_2$.
Finally,
\begin{equation}\label{ea3-12}
\ds  t_m=\frac{t}{(1+V)}.
\end{equation}

To simulate the maximum $M$ conditional on being in an interval $[M_1,M_2]$, we use the symmetry of Brownian bridge by simulating a minimum $m_0$, conditional on being in the interval $[-M_2,-M_1]$, and its location $\ds t_{m_0}$ and making
\begin{equation}\label{ea3-13}
\ds   M=-m_0\;\;\mbox{and}\;\;t_M=t-t_{m_0}.
\end{equation}

Let $\mathbb{B}\mathbb{B}_{D_I}$ be the measure of the layered Brownian bridge, we accept the proposal w.p.
\begin{equation}\label{ea3-14}
\ds  \frac{d\mathbb{B}\mathbb{B}_{D_I}}{d\mathbb{P}_{D_I}}(W)\propto\frac{\mathbb{I}[D_I]}{1+\mathbb{I}[\overline{D}_I\bigcap \underline{D}_I]}\leq1.
\end{equation}
In order to compute this probability we need to simulate the two Bernoulli r.v.'s in (\ref{ea3-14}). The first one (in the numerator) is equal to 1 w.p.
\begin{eqnarray}\label{ea3-15}
\ds  \delta(t_m,0,-m,b_I-m)&\times&\delta(t-t_m,0,-m,b_I-m) \mbox{ or } \\ \delta(t_M,0,M,M+b_I)&\times&\delta(t-t_M,0,M,M+b_I)).
\end{eqnarray}
and the second one (in the denominator) is 1 w.p. 1 if $I=1$ and is 0 w.p.
\begin{eqnarray}\label{ea3-16}
\ds  \delta(t_m,0,-m,b_{I-1}-m;b_I-m)&\times&\delta(t-t_m,0,-m,b_{I-1}-m;b_I-m) \mbox{ or } \\ \delta(t_M,0,M,M+b_{I-1};M+b_I)&\times&\delta(t-t_M,0,M,M+b_{I-1};M+b_I),
\end{eqnarray}
if $I>1$, where
\begin{eqnarray}
\ds  \delta\left(s,0,u_2,K\right)&=&1-\frac{1}{u_2}\sum_{j=1}^{\infty}\left\{\zeta_j(s,u_2,K)-\xi_j(s,u_2,K)\right\}\label{ea3-17a}\\
\delta\left(s,0,u_2,K;L\right)&=&\frac{u_2-\sum_{j=1}^{\infty}\left\{\zeta_j(s,u_2,K)-\xi_j(s,u_2,K)\right\}}{u_2-\sum_{j=1}^{\infty}\left\{\zeta_j(s,u_2,L)-\xi_j(s,u_2,L)\right\}}\label{ea3-17b}\\
\xi_j(s,u_2,K)&=&(2Kj+u_2)\exp\{-2Kj(Kj+u_2)/s\}\label{ea3-17c}\\
\zeta_j(s,u_2,K)&=&\xi_j(s,-u_2,K).\label{ea3-17d}
\end{eqnarray}
The alternating series method can be used for the events involving the functions $\delta$ if $3K^2>s$. Otherwise, it can be used after the first $\lceil\sqrt{t+K^2}/(2K)\rceil$ pairs $S_{2j-1}$ and $S_{2j}$ have been computed. This follows from the result in \citet{pollock} (Corollary 5) that states that the sequence of bounds for the probabilities in (\ref{ea3-15}) and (\ref{ea3-16}) are Cauchy sequences after the first $\lceil\sqrt{t+K^2}/(2K)\rceil$ pairs are computed.

Finally note that the acceptance probability is 0 if the Bernoulli r.v. in the numerator is 0. If this r.v. is 1, the acceptance probability is 1 if the Bernoulli r.v. in the denominator is 0 and 1/2 if it is 1.

The algorithm outputs the following variables under the measure $\mathbb{B}\mathbb{B}_{D_I}$: $\ds (m^*,t^*)=(m,t_m)\mbox{ or }(M,t_M)$; $\mathbb{I}[\overline{D}_I\bigcap \underline{D}_I]$. We choose to keep only $(m^*,t^*)$ which is an exact draw from $\ds \pi_{\mathbb{B}\mathbb{B}_{D_I}}(m^*,t^*)$ - the density of $(m^*,t^*)$ under $\mathbb{B}\mathbb{B}_{D_I}$.

At this stage of the layered Brownian bridge algorithm we have upper and lower bounds for the standard Brownian bridge $W$, which are given by\\
$\left\{
  \begin{array}{ll}
    \dot{z}=m \mbox{ and } \dot{u}=b_I, & \mbox{ if the proposal chooses } BB_{\underline{M}_I}; \\
    \dot{z}=-b_I \mbox{ and } \dot{u}=M, & \mbox{ if the proposal chooses } BB_{\overline{M}_I}.
  \end{array}
\right.$

We now proceed to the simulation of the bridge points given the information we have. We represent this by $\mathcal{I}_{m^*,I}$, which is either the minimum and its location and an upper bound for $W$ or the maximum and its location and a lower bound for $W$. The bridge points are simulated via rejection sampling by proposing from $\mathbb{B}\mathbb{B}|(m^*,t^*)$ and accepting the proposal if $\mathbb{I}[D_I]=1$.

The law $\mathbb{B}\mathbb{B}|(m^*,t^*)$ is a three-dimensional Bessel bridge and is constructed as follows. Firstly, denote by  $\ds R(\delta)=\{R_s(\delta);\;0\leq s\leq 1\}$ a three-dimensional Bessel bridge of unit length from 0 to $\delta\leq0$, i.e.
\begin{equation}\label{ea3-19}
\ds   R_s(\delta)=\sqrt{(\delta s+W_{s}^{(1)})^2+(W_{s}^{(2)})^2+(W_{s}^{(3)})^2},\;\;s\in[0,1],
\end{equation}
where the $W_{s}^{(\cdot)}$'s are three independent standard Brownian bridges in $[0,1]$. Then, defining $W^*$ to be $W|(m,t_m)$, we have that
\begin{eqnarray}\label{ea3-20}
\ds     \{W_{s}^*;\;0\leq s\leq t_m\}&=&\sqrt{t_m}R_{(t_m-s)/(t_m)}(\delta_1)+m,\\
\ds     \{W_{s}^*;\;t_m\leq s\leq t\}&=&\sqrt{t-t_m}R_{(s-t_m)/(t-t_m)}(\delta_2)+m,
\end{eqnarray}
where $\ds\delta_1=\frac{-m}{\sqrt{t_m}}$ and $\ds\delta_2=\frac{-m}{\sqrt{t-t_m}}$.

If we now define $W^*$ to be $W|(M,t_M)$, we have that
\begin{eqnarray}\label{ea3-21}
\ds   \{W_{s}^*;\;0\leq s\leq t_M\}&=&M-\sqrt{t_M}R_{(t_M-s)/(t_M)}(\delta_1),\\
\ds   \{W_{s}^*;\;t_M\leq s\leq t\}&=&M-\sqrt{t-t_M}R_{(s-t_M)/(t-t_M)}(\delta_2),
\end{eqnarray}
where $\ds \delta_1=\frac{M}{\sqrt{t_M}}$ and $\ds \delta_2=\frac{M}{\sqrt{t-t_M}}$.

This construction implies that $\mathbb{B}\mathbb{B}|\mathcal{I}_{m^*,I}$ is a Markov process and, therefore, the proposed rejection sampling algorithm can be performed interval-wise w.r.t. the sub-intervals defined by the bridge points already simulated. Furthermore, in order to simulate the proposal, all the points from the three standard Brownian bridges $W_{s}^{(\cdot)}$ need to be stored along the way.

Finally, once the proposal in simulated in a given sub-interval, the acceptance indicator is simulated via the alternating series method. Suppose that $\mathbb{B}\mathbb{B}|\mathcal{I}_{m^*,I}$ is to be simulated at locations $(s_1,\ldots,s_n)$ in a sub-interval $(s_0,s_{n+1})$. Note that $s_0$ and $s_{n+1}$ may be one of the extremes of $(0,t)$ or even $t^*$. We have that
\begin{equation}\label{ea3-18}
\ds P(D_I)=\prod_{k=1}^{n+1}\gamma\left(s_k-s_{k-1};|m^*-W_{s_{k-1}}^*|,|m^*-W_{s_k}^*|,|m^*|+b_I\right),
\end{equation}
where
$$\ds  \delta\left(s,u_1,u_2,K\right)=\frac{\gamma(s,u_1-K/2,u_2-K/2,K/2)}{1-\exp\{-2u_1u_2/s\}}.$$

The choice of the sequence $b_1,b_2,\ldots$, specially $b_1$, is of great concern. On one hand, if $b_1$ is a relatively high value, we will get $I=1$ more often, which improves the computational time for sampling from $W$. On the other hand, a high value of $b_1$ leads to conservative bounds for the function $\phi$ in the two steps of the Barker's MCMC which in turn leads to an inefficient two-coin algorithm. For an interval of length $t$, we suggest $\sqrt{t}/2 \leq b_1 \leq \sqrt{t}$.

\section*{Appendix F - Obtaining efficient lower bounds $\dot{a}_{i,j}(s;\theta)$}

Let $\dot{X}_{i,j}$ be the standard BB in $s\in(\tau_{i,j-1},\tau_{i,j})$ (simulated from $\tilde{\mathbb{D}}$) and define $\dot{z}_{i,j}$ and $\dot{u}_{i,j}$ to be, respectively, $\dot{z}$ and $\dot{u}$, as defined in Appendix E when applying the layered Brownian bridge algorithm to the interval $(\tau_{i,j-1},\tau_{i,j})$. This means that $\dot{z}_{i,j}$ and $\dot{u}_{i,j}$ are lower and upper bounds for $\dot{X}_{i,j}$. Now define ${x}_{i,j}:={x}_{i,j}(\theta)=\min\{X_{i,j-1},X_{i,j-}\}$, ${y}_{i,j}:={y}_{i,j}(\theta)=\max\{X_{i,j-1},X_{i,j-}\}$,
\begin{equation}\label{layer1}
\ds z_{i,j}(s;\theta)=\left\{
            \begin{array}{ll}
              \frac{(y_{i,j}-x_{i,j})}{\Delta \tau_{i,j}}(s-\tau_{i,j-1})+({x}_{i,j}+\dot{z}_{i,j}), & \mbox{if }{x}_{i,j}=X_{i,j-1} \\
             -\frac{(y_{i,j}-x_{i,j})}{\Delta \tau_{i,j}}(s-\tau_{i,j})  +({x}_{i,j}+\dot{z}_{i,j}), & \mbox{if }{x}_{i,j}=X_{i,j},
            \end{array}
          \right.
\end{equation}
and
\begin{equation}\label{layer2}
\ds u_{i,j}(s;\theta)=\left\{
            \begin{array}{ll}
              \frac{(y_{i,j}-x_{i,j})}{\Delta \tau_{i,j}}(s-\tau_{i,j-1})+({x}_{i,j}+\dot{u}_{i,j}), & \mbox{if }{x}_{i,j}=X_{i,j-1} \\
             -\frac{(y_{i,j}-x_{i,j})}{\Delta \tau_{i,j}}(s-\tau_{i,j})  +({x}_{i,j}+\dot{u}_{i,j}), & \mbox{if }{x}_{i,j}=X_{i,j},
            \end{array}
          \right.
\end{equation}
where $\Delta \tau_{i,j}=\tau_{i,j}-\tau_{i,j-1}$. This implies that $\ds X_s(\theta)=\varphi(\dot{X}_s;\theta)\in\left[z_{i,j}(s;\theta),u_{i,j}(s;\theta)\right]=:H_{i,j}(s;\theta)$, $\forall\; s\in[\tau_{i,j-1},\tau_{i,j})$.

For $\dot{\phi}$ as defined in equation (11) in Subsection 3.1.2 of the paper, define
\begin{equation}\label{optbnd1}
z_{\dot{\phi}_{i,j}}(s;\theta)=\inf_{\varphi(u;\theta)\in H_{i,j}(s)}\dot{\phi}(u,s;\theta)\;\; \mbox{and}\;\;Z_{i,j}(\theta)=\inf_{s\in[t_{i_{j-1}},t_{i_j})}z_{\dot{\phi}_{i,j}}(s;\theta),
\end{equation}
\begin{equation}\label{optbnd2}
u_{\dot{\phi}_{i,j}}(s;\theta)=\sup_{\varphi(u;\theta)\in H_{i,j}(s)}\dot{\phi}(u,s;\theta)\;\; \mbox{and}\;\;U_{i,j}(\theta)=\sup_{s\in[t_{i_{j-1}},t_{i_j})}u_{\dot{\phi}_{i,j}}(s;\theta),
\end{equation}
If we can compute $\ds\int_{t_{i_{j-1}}}^{t_{i_j}}z_{\dot{\phi}_{i,j}}(\dot{X}_s,s;\theta)ds$, we make $\ds \dot{a}_{i,j}(s;\theta)=z_{\dot{\phi}_{i,j}}(s;\theta)$, for $s\in[t_{i_{j-1}},t_{i_j})$, otherwise $\ds \dot{a}_{i,j}(s;\theta)=Z_{i,j}(\theta)$. Finally, we define $\dot{r}_{i,j}(\theta)=U_{i,j}(\theta)-Z_{i,j}(\theta)$.

Functions $z_{i,j}(s;\theta)$ and $u_{i,j}(s;\theta)$ in (\ref{layer1}) and (\ref{layer2}) can be replaced by the constants and more conservative bounds (but simpler to do the computation) $z_{i,j}(\theta)= x_{i,j} + \dot{z}_{i,j}$ and $u_{i,j}(\theta)=y_{i,j} + \dot{u}_{i,j}$.

\section*{Appendix G - Auxiliary results}

Consider the definitions of $\theta$, $\mathbb{P}$, $\mathbb{Q}$, $\mathbb{L}$, $\mathbb{W}_{0,0}$, $\mathbf{v}$, $\mathbb{J}_{0,i}$, $\mathbf{J}^{(i)}$, $N_i$, $X_{J}^{(i)}$, $\ddot{X}^{(i)}$, $\tau_{i,j}$, for $i=1,\ldots,n$ and $j=1,\ldots,N_i$, as presented in the paper.

Now define the product measures $\ds \mathbb{H}_{k,i}:=\mathbb{J}_{0,i}\otimes\mathbb{L}^k\otimes\mathbb{W}_{0,0}^{k+1}$, for $k=0,1,2,\ldots$, and $\ds \mathbb{H}_i=\sum_{k=0}^{\infty}\mathbb{H}_{k,i}$, for $i=1,\ldots,n$.

We state the following proposition which shall be useful in the proof of Lemma 1 from the paper.
\begin{proposition}\label{prop_sup}
\begin{equation}\label{ies5}
\ds \frac{d\mathbb{P}}{d\mathbb{H}_i}(\mathbf{J}^{(i)},X_{J}^{(i)},\ddot{X}^{(i)}|\mathbf{v},\theta)= \frac{d\mathbb{P}}{d\mathbb{Q}}(\mathbf{J}^{(i)},X_{J}^{(i)},\ddot{X}^{(i)}|\mathbf{v},\theta) \frac{d\mathbb{Q}}{d\mathbb{H}_i}(X_{J}^{(i)}|\mathbf{J}^{(i)},\mathbf{v},\theta).
\end{equation}
\end{proposition}

\begin{thma}
We have that,
\begin{equation*}
\ds \frac{d\mathbb{P}}{d\mathbb{H}_i}(\mathbf{J}^{(i)},X_{J}^{(i)},\ddot{X}^{(i)}|\mathbf{v},\theta)=
  \frac{d\mathbb{P}}{d\mathbb{Q}}(\mathbf{J}^{(i)},X_{J}^{(i)},\ddot{X}^{(i)}|\mathbf{v},\theta) \frac{d\mathbb{Q}}{d\mathbb{H}_{i}}(\mathbf{J}^{(i)},X_{J}^{(i)},\ddot{X}^{(i)}|\mathbf{v},\theta).
\end{equation*}

Then, defining $\ddot{X}^{(i,j)}$ to be $\ddot{X}^{(i)}$ in $(\tau_{i,j-1},\tau_{i,j})$, we have
\begin{eqnarray*}
\ds \frac{d\mathbb{Q}}{d\mathbb{H}_{i}}(\mathbf{J}^{(i)},X_{J}^{(i)},\ddot{X}^{(i)}|\mathbf{v},\theta)
&=& \frac{d\mathbb{Q}}{d\mathbb{H}_{i}}(\mathbf{J}^{(i)}|\mathbf{v},\theta)\frac{d\mathbb{Q}}{d\mathbb{H}_i}(X_{J}^{(i)}|\mathbf{J}^{(i)},\mathbf{v},\theta)\frac{d\mathbb{Q}}{d\mathbb{H}_{i}}(\ddot{X}^{(i)}|\mathbf{J}^{(i)},X_{J}^{(i)},\mathbf{v},\theta) \\
&=&\frac{d\mathbb{Q}}{d\mathbb{H}_i}(X_{J}^{(i)}|\mathbf{J}^{(i)},\mathbf{v},\theta).
\end{eqnarray*}

\end{thma}

\newpage

\section*{Appendix H - Further results from simulations}

For the two models considered in Subsections 4.1 and 4.2, we present the results for two other data sets.

\begin{table}[!h]
\centering
\fbox{\scriptsize
\begin{tabular}{c|ccccc}
        & $\delta$ & $\sigma^2$ & $\lambda$ & $\mu$ & $\tau^2$ \\ \hline
        & -0.072 & 0.851 & 0.175 & 1.610 & 0.201 \\ \hline
 real   & 0  & 1  & 0.1  & 2  & 0.1225
\end{tabular}}
\caption{\label{tabsim1}Last iteration values for the MCEM algorithm applied to the second data set from the model in Subsection 4.1.}
\end{table}

\begin{table}[!h]
\centering
\fbox{\scriptsize
\begin{tabular}{c|ccccc}
              & $\delta$ & $\sigma^2$ & $\lambda$ & $\mu$ & $\tau^2$ \\ \hline
  Mean        & -0.138 & 0.827 & 0.260 & 1.441 & 0.254 \\
  Median      & -0.136 & 0.826 & 0.250 & 1.415 & 0.244 \\
  Mode        & -0.142 & 0.830 & 0.203 & 1.402 & 0.235 \\
  St. Dev.    & 0.106  & 0.069  & 0.103 & 0.232 & 0.121 \\ \hline
  real        & 0 & 1 & 0.1 & 2 & 0.1225
\end{tabular}}
\caption{\label{tabsim2}Posterior statistics from the MCMC output for the second data set from the model in Subsection 4.1.}
\end{table}

\begin{table}[!h]
\centering
\fbox{\scriptsize
\begin{tabular}{c|ccccc}
        & $\delta$ & $\sigma^2$ & $\lambda$ & $\mu$ & $\tau^2$ \\ \hline
        & -0.086 & 0.956 & 0.147 & 1.973 & 0.170 \\ \hline
 real   & 0  & 1  & 0.1  & 2  & 0.1225
\end{tabular}}
\caption{\label{tabsim1}Last iteration values for the MCEM algorithm applied to the third data set from the model in Subsection 4.1.}
\end{table}

\begin{table}[!h]
\centering
\fbox{\scriptsize
\begin{tabular}{c|ccccc}
              & $\delta$ & $\sigma^2$ & $\lambda$ & $\mu$ & $\tau^2$ \\ \hline
  Mean        & -0.103 & 0.960 & 0.172 & 1.863 & 0.325 \\
  Median      & -0.102 & 0.959 & 0.164 & 1.861 & 0.307 \\
  Mode        & -0.098 & 0.953 & 0.153 & 1.895 & 0.323 \\
  St. Dev.    &  0.071 & 0.063 & 0.055 & 0.268 & 0.194 \\ \hline
  real        & 0 & 1 & 0.1 & 2 & 0.1225
\end{tabular}}
\caption{\label{tabsim2}Posterior statistics from the MCMC output for the third data set from the model in Subsection 4.1.}
\end{table}

\begin{table}[!h]
\centering
\fbox{\scriptsize
\begin{tabular}{c|ccccc}

          & $\rho$ & $\mu$ & $\lambda$ & $\theta$  \\ \hline

          & 0.967  &  -0.051 &  0.173    &  0.831 \\ \hline
 real     & 1      & 0      &   0.07    & 1
\end{tabular}}
\caption{\label{tabsim4}Last iteration values for the MCEM algorithm applied to the second data set from the model in Subsection 4.2.}
\end{table}

\begin{table}[!h]
\centering
\fbox{\scriptsize
\begin{tabular}{c|c|c|c|c}
  & $\rho$ & $\mu$ & $\lambda$ & $\theta$ \\ \hline
  Mean          & 1.019 & -0.055 & 0.086 & 0.558 \\
  Median        & 1.020 & -0.054 & 0.083 & 0.549 \\
  Mode          & 1.034 & -0.056 & 0.078 & 0.535 \\
  St. Dev.      & 0.071 &  0.053 & 0.029 & 0.118 \\
  real          & 1 & 0 & 0.07 & 1
\end{tabular}}
\caption{\label{tabsim6}Posterior statistics from the MCMC output for the second data set from the model in Subsection 4.2.}
\end{table}

\begin{table}[!h]
\centering
\fbox{\scriptsize
\begin{tabular}{c|ccccc}

          & $\rho$ & $\mu$ & $\lambda$ & $\theta$  \\ \hline
          & 0.926  &  -0.054 &  0.145    &  0.767 \\ \hline
 real     & 1      & 0      &   0.07    & 1
\end{tabular}}
\caption{\label{tabsim4}Last iteration values for the MCEM algorithm applied to the third data set from the model in Subsection 4.2.}
\end{table}

\begin{table}[!h]
\centering
\fbox{\scriptsize
\begin{tabular}{c|c|c|c|c}
  & $\rho$ & $\mu$ & $\lambda$ & $\theta$ \\ \hline
  Mean          & 0.854 & 0.113 & 0.043 & 0.934 \\
  Median        & 0.855 & 0.114 & 0.038 & 0.900 \\
  Mode          & 0.854 & 0.112 & 0.032 & 0.865 \\
  St. Dev.      & 0.069 & 0.060 & 0.025 & 0.278 \\
  real          & 1 & 0 & 0.07 & 1
\end{tabular}}
\caption{\label{tabsim6}Posterior statistics from the MCMC output for the third data set from the model in Subsection 4.2.}
\end{table}

\end{document}